*Review*

# Quantum Dots as Functional Nanosystems for Enhanced Biomedical Applications

Pronama Biswas[1,*], Asmita Saha[1], Bhoomika Sridhar[1], Anwesha Patel[1], Belaguppa Manjunath Ashwin Desai[2]

[1]School of Basic and Applied Sciences, Dayananda Sagar University, Shavige Malleshwara Hills, 1st Stage, Kumaraswamy Layout, Bengaluru, KA 560078, India
[2]School of Engineering, Dayananda Sagar University, Bengaluru, KA 560068, India

*Correspondence: pronamabiswas@gmail.com; pronama-sbas@dsu.edu.in (Pronama Biswas)

## Abstract

Quantum dots (QDs) have emerged as promising nanomaterials with unique optical and physical properties, making them highly attractive for various applications in biomedicine. This review provides a comprehensive overview of the types, modes of synthesis, characterization, applications, and recent advances of QDs in the field of biomedicine, with a primary focus on bioimaging, drug delivery, and biosensors. The unique properties of QDs, such as tunable emission spectra, long-term photostability, high quantum yield, and targeted drug delivery, hold tremendous promise for advancing diagnostics, therapeutics, and imaging techniques in biomedical research. However, several significant hurdles remain before their full potential in the biomedical field, like bioaccumulation, toxicity, and short-term stability. Addressing these hurdles is essential to effectively incorporate QDs into clinical use and enhance their influence on healthcare outcomes. Furthermore, the review conducts a critical analysis of potential QD toxicity and explores recent progress in strategies and methods to mitigate these adverse effects, such as surface modification, surface coatings, and encapsulation. By thoroughly examining current research and recent advancements, this comprehensive review offers invaluable insights into both the future possibilities and the challenges that lie ahead in fully harnessing the potential of QDs in the field of biomedicine, promising a revolution in the landscape of medical diagnostics, therapies, and imaging technologies.

## Introduction

Quantum dots (QDs), first discovered by Alexie Ekimov and Onushenko in 1981 within a glass matrix [1], are nanoscale semiconductor particles ranging in size from 1 to 10 nm. They have garnered significant attention and found extensive applications in diverse fields. QDs have unique optical properties, including remarkable fluorescence with tunable emission spectra and high quantum yields. These properties make QDs ideal for precisely controlling emitted light [2]. Consequently, they are used in optoelectronic devices like quantum light-emitting diodes (QLEDs) and solar cells. QDs demonstrate exceptional electrochemical properties, including high charge mobility and stability. These properties enable precise control and efficient charge utilization, making QDs highly suitable for applications in sensors and batteries.

Furthermore, the small size of QDs allows them to penetrate biological membranes easily, opening possibilities for applications in cellular imaging and targeted drug delivery systems. QDs have demonstrated outstanding potential in bioimaging techniques, enabling high-resolution imaging of cells, tissues, and even individual biomolecules [3]. They can be exploited as fluorescent labels for cellular tracking, diagnostics, and biomolecular sensing, enabling diverse applications, including cancer detection, bacterial and viral infection monitoring, immunoassays, and *in-vivo* cell tracking [4]. QDs can target specific cells, like antibodies, and can be employed in multi-analyte detection [5]. Their size variability and ability to modify surface coatings enhance their versatility, making them vital in the biomedical domain.

However, it is crucial to address potential challenges associated with using QDs, including long-term stability, potential toxicity, and environmental impact. The small size and reactivity of QDs can result in their bioaccumulation in the host system, potentially causing inflammation and other adverse effects [6]. Researchers are actively improving the biocompatibility of QDs through surface modification techniques and exploring alternative materials to mitigate any potential adverse effects [7]. Continued research efforts are underway to enhance the functionality of QDs and ensure their safe and responsible utilization in various applications.

The primary focus of this review is to provide a comprehensive overview of the types, synthesis, characterization, applications, and recent advances of QDs in biomedicine, including bioimaging, drug delivery, and biosensors. Furthermore, the potential toxicity of QDs is critically examined, and advances in the current strategies and approaches to mitigate their adverse effects are discussed. By evaluating the current state of research and advancements in QDs, this review aims to provide insights into the future directions and challenges in harnessing the full potential of QDs while ensuring their safe and responsible utilization.

**Types of QDS**

QDs can be classified into various types based on criteria, including elemental composition, properties, size, shape, and structure (Table 1, Ref. [7-15, 23, 26-28, 32, 34,36]), each carefully designed to cater to a specific application area.

**Table 1. Classification of QDs, their examples, properties, and applications.**

| Criteria | Types | Examples | Properties | Applications | Ref. |
|---|---|---|---|---|---|
| Elemental Composition | I B-VI A | CuInS$_2$ | Broad emission absorption spectrum, high photoluminescence | Optoelectronic devices, biological imaging | [8] |
| | I B-VII A | AgBr | Plasmonic property, stability | Optical devices, sensors | [8] |
| | II B-VI A | ZnS | Versatile and reliable fluorescence emission | Solar cells, photovoltaics, biological imaging | [9] |
| | III A-V A | BN | Higher optically stability, low quantum yield | Optoelectronic devices, quantum computing | [9] |
| | IV A-VI A | Carbon QDs | Biocompatibility, smaller energy gaps | Photovoltaics, sensors, nanoelectronics | [10] |
| | V A | Nitrogen QDs | Biocompatibility, high stability | Sensors, optoelectronic devices, computing | [8] |
| | I B-III A-VI A | CuInS$_2$ | Longer fluorescent lifetime, low biological toxicity | Solar cells, light-emitting devices | [8] |
| | Pdots | Poly(ethyleneimine) dots | Higher brightness, stability, and low toxicity | Biological imaging, biosensors | [11] |
| | TMDCs | TMDC-WS$_2$ | Tunable emission spectrum, biocompatibility | Optoelectronic devices, cancer therapy | [12] |
| | MXene | - | Excellent electrical conductivity | Energy storage devices, sensors | [13] |
| | Perovskite | Formamidinium PbBr3 PQDs | Room-temperature synthesis, good photo-luminescence | LEDs, photodetectors | [14] |
| Size | Small- QDs | CdSe QDs | Quantum confinement effects, larger bandgap | Biological imaging, optoelectronic devices | [15] |
| | Large-QDs | CdTe QDs | Near-infrared light emission, smaller bandgap | LEDs | [23] |
| Shape | Spherical QDs | CdSeS QDs | Symmetrical emission and isotropic emission | LEDs, biological imaging | [26] |
| | Nanorods | CdTe QDs | Anisotropic emission and directional emission | Displays, waveguides, optoelectronics | [27] |
| Structure | Core-Shell QDs | CdSe/ZnS QDs | Shell enhances stability and quantum yield, low cytotoxicity | Optoelectronic devices, biological imaging | [28] |
| | Janus QDs | Triangular Janus MoSSe QDs | Asymmetric functionality, tailored interfacial interactions | Catalysis, sensing, biological imaging | [32] |
| | Alloyed QDs | InGaAs QDs | Tunable optical spectra, tunable bandgap | Pollutant degradation, energy storage | [34] |
| | Doped QDs | Mn-doped ZnSe QDs | Broad and strong absorption, tuneable optical properties | Optoelectronic devices | [7,36] |

QDs, Quantum dots; TMDCs, Transition Metal Dichalcogenide QDs; PQDs, peptide-paramagnetic QDs; LEDs, light-emitting diodes; Pdots, Polymer dots; ZnS, Zinc Sulphide; InGaAs, Indium Gallium Arsenide; CdSe, Cadmium Selenium QD.

*Elemental Composition*

Based on the composition of the elements or materials making up QD, QDs can be classified into 12 types: I B-VI A, I B-VII A, II B-VI A, III A-V A, IV A-VI A, IV A, V A, I B-III A-VI A, Polymer dots (Pdots), Transition Metal Dichalcogenide QDs (TMDCs), MXene QDs (MXQDs), and Perovskite QDs. Each type possesses distinct compositions, giving them specific properties and making them useful in diverse fields (Table 1). For example, the I B-VI A type QDs comprise elements from Group I B (copper/ silver/ gold) and Group VI A (carbon/ silicon/ germanium) of the periodic table. Type I B-VII A consists of elements such as copper, silver, gold, fluorine, chlorine, and bromine, which have specific properties that benefit photonics and sensing [8]. Type II B-VI A QDs, made of zinc/ cadmium/ mercury and sulfur/ selenium/ tellurium, have diverse applications in displays, solar cells, and biological imaging [9]. Type III A-V A, consisting of boron/ aluminium/ gallium/ indium and nitrogen/ phosphorus/ antimony, is used in optoelectronics and quantum computing [9]. Type IV A-VI A, which include carbon/ silicon/ germanium and sulfur/ selenium/ tellurium, are versatile and are used in solar cells, sensors, and nanoelectronics [10]. Type V A QDs containing nitrogen/ phosphorous/ arsenic/ antimony have distinctive properties suitable for sensors, optoelectronics, and quantum computing [8]. On the other hand, type I B-III A-VI A QDs are composed of a combination of copper/ silver/ gold, boron/ aluminium/ indium/ gallium, and sulfur/ selenium/ tellurium. For example, copper indium sulfide is an excellent material for solar harvesting and light-emitting applications [8]. Pdots are QDs composed of organic polymers like polyethylene glycol (PEG) and poly-acrylamide and are suitable for applications in bioimaging and biosensing [11]. TMDCs are composed of layered materials containing transition metals (like molybdenum) and chalcogen elements (like sulfur) and have excellent optical and electronic properties [12]. MXQDs are two-dimensional nanolayered QDs composed of transition metal carbides, nitrides, or carbonitrides known for their excellent electrical conductivity and are useful in energy storage devices and sensors [13]. Perovskite QDs, made of hybrid organic-inorganic materials with a perovskite crystal structure ($ABX_3$), have excellent optoelectronic properties and are used in light-emitting diodes (LEDs) and photodetectors [14].

*Properties of QDs*

Recent advances in QD nanotechnology demonstrate promise in both electrical and biological applications owing to their adjustable optoelectronic features, high photostability, quantum yield, and significant molar extinction coefficients, which uniquely position QDs to encompass all these advantages, setting them apart from organic dyes and rendering them highly suitable for diverse imaging and biosensing applications [15,16]. One suitable example of this is Carbon QDs (CQDs), which have garnered significant interest over recent years as a viable alternative to semiconductor QDs because of their biocompatibility, stability, high photoluminescence (PL), and low cost [17]. QDs are exceptionally brilliant and reliable instruments for use in fluorescence applications. They have been successfully utilized in animal models, such as sentinel lymph node mapping and tumor diagnosis. Whole-body and microscopic fluorescence methods enabled researchers to detect QD fluorescence in mice for two years post-injection. Two-photon spectrum imaging confirmed their presence but revealed considerably blue-shifted emission peaks, indicating that systemically administered QDs can persist *in vivo* for up to two years, although with notable blue-shifted emission [18]. QDs offer versatile applications by targeting specific cells or proteins with peptides, antibodies, or ligands, enabling the investigation of a target protein or cell behavior. Their potential is evident in receptor-based targeting. Functionalized QDs (f-QDs) are efficient, safe, nano-sized systems for delivering various bio-actives. Surface-modified fluorescent carbon QDs, acting as targeting ligands, have garnered interest for cellular targeting with enhanced specificity, particularly in ongoing research for cancer therapy [19]. Another property is high quantum yield, which refers to the efficiency of a material or substance in converting absorbed photons into emitted photons, enhancing their properties [20]. QD solutions exhibit distinct colors when exposed to Ultraviolet (UV) light due to variations in particle sizes. The efficiency of energy transfer properties is heightened by surface area effects, leading to the well-known characteristic of QD having a high quantum yield [21]. Researchers identified constant emission wavelengths in PL spectra, with carbon QD at around 530 nm and titanium dioxide ($TiO_2$) carbon dots at 496 nm, independent of reaction circumstances. Tuning operating settings allows carbon dots to obtain a maximum quantum yield of 18%, whereas $TiO_2$-carbon dots achieve an astonishing 68%. N-doped $TiO_2$ nanoparticles as nucleation seeds improve quantum yield even more, achieving an even higher value of 85.9% [17].

*Size of QDs*

QDs can also be classified based on size, which is crucial in determining their electronic and optical properties. Size-based classifications categorize them into small- and large-sized QDs. Small-sized QDs (1–4 nm) exhibit unique characteristics due to their small size, resulting in quantum confinement effects. They have larger bandgaps, requiring more energy to excite electrons to higher energy states [22]. As a result, they emit light at shorter wavelengths. Large-sized QDs (5–10 nm) have smaller bandgaps and require less energy to excite electrons. Consequently, they emit light at longer wavelengths, which allows the absorption and emission of lower-energy photons [23]. Small QDs are utilized for imaging, specifically sub-cellular structures, as their emissions have better tissue penetration and reduced scattering effects, making them valuable for *in-vivo* imaging [24]. On the other hand, large-sized QDs can be functionalized with tumor-targeting ligands or antibodies, enabling sensitive and specific cancer detection [25].

*Shape of QDs*

The shape of QDs plays a crucial role in determining their properties. Shape-based classifications provide insights into the diverse geometries of QDs, which can significantly impact their optical and electronic characteristics. Two common shape-based classifications are spherical QDs and nanorods. Spherical QDs exhibit symmetrical properties due to their uniform shape, resulting in isotropic emission characteristics. The spherical symmetry of these QDs allows for light emission in all directions equally. The uniform emission properties make spherical QDs particularly useful for applications requiring homogeneous light emission, such as LEDs and biological imaging [26]. Nanorods, on the other hand, have an elongated rod-like shape. They possess anisotropic properties that vary depending on the orientation. Nanorods exhibit directional emission characteristics; light is preferentially oriented along the elongated axis of the rod. The anisotropic nature of nanorods opens opportunities for applications where controlled emission in specific directions is desired, such as in displays, waveguides, and optoelectronic devices [27].

*Structure of QDs*

Structure-based classifications offer insights into the specific structural features of QDs, providing a deeper understanding of their behavior and potential applications. The common structure-based classification includes core-shell, Janus, alloyed, and doped QDs (Fig. 1).

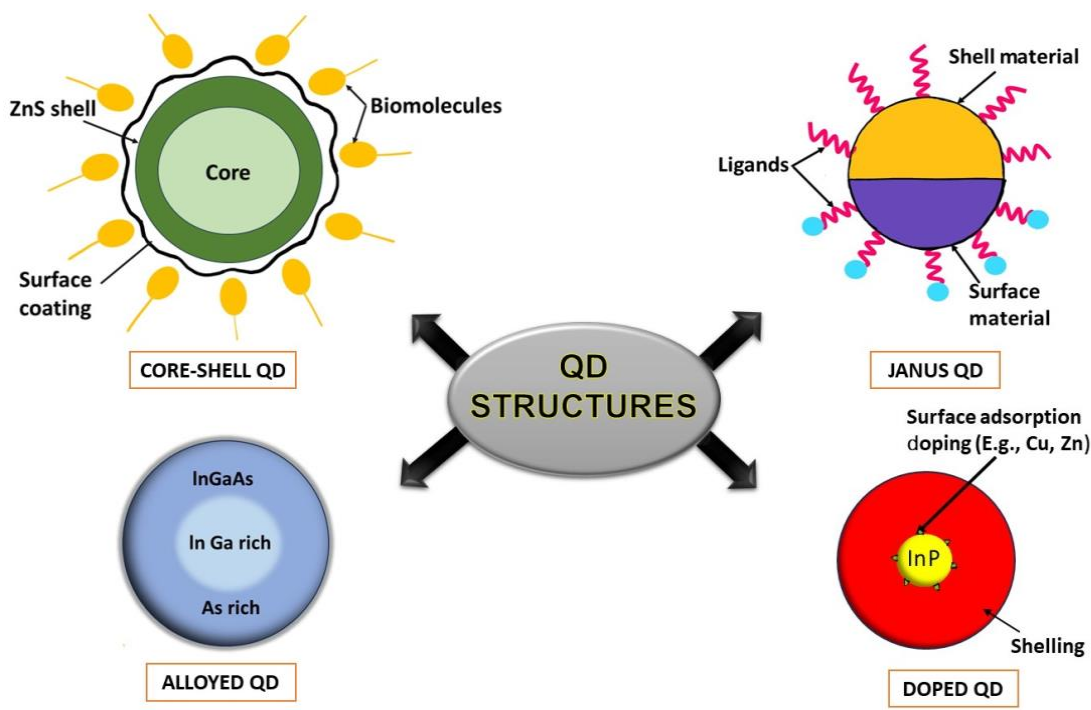

**Fig. 1. Classification of Quantum Dots (QDs) based on structure.** Core-Shell QD: covering the core with a semiconductor material with a larger band gap, like ZnS, can increase stability and quantum yield while simultaneously neutralizing the toxicity of the core by preventing photo-oxidative conditions like Ultraviolet (UV) and air from exposing reactive $Cd^{+2}$ and $Te^{+2}$ ions. Alloyed QD: Indium is spread across InGaAs QDs; these QDs span a broad infrared optical range. Janus QD: composed of three parts, i.e., shell material, ligand, and surface material, where the overall structure of Janus is asymmetric, providing selective interaction in solution. Doped QD: The typical III-V group semiconductor nanocrystals like InP QDs have a large excitonic Bohr's radius and strong carrier mobility. QD, Quantum dot; ZnS, Zinc Sulphide; InGaAs, Indium Gallium Arsenide; InP, Indium Phosphide;

Cu, Copper; Zn, Zinc. The diagram was made using Microsoft® PowerPoint® (Version 2310 Build 16.0.16924.20054).

Core-shell QDs are characterized by a core material surrounded by a shell material. The core provides the desired electronic and optical properties, while the shell enhances the stability of the QDs and improves their optical properties, such as quantum yield and PL efficiency. By carefully selecting the core and shell materials, core-shell QDs with enhanced stability, reduced sensitivity to external factors, and improved photostability can be engineered. Examples of core-shell QDs include Cadmium Selenium QD (CdSe)/ZnS and CdTe/CdSe [28]. The core of these QDs can be highly toxic and unstable due to its high surface area-to-volume ratio. Capping the core with a semiconductor material of a higher band gap (like ZnS) can reduce their toxicity, increase stability, and increase performance at room temperatures [29]. Along with capping, biomolecules like antibodies, peptides, and nucleic acids can be attached (Fig. 1) to the surface coating, which gives them the ability to bioconjugate [30]. The utilization of core-shell particles in biomedical applications has been made possible by their improved compatibility, potential reduction in cytotoxicity, potential for conjugating bioactive compounds, and higher stability resulting from the protective shell. Folic acid (FA) enhances the capability for cancer cell entry via receptor-mediated endocytosis when introduced to the core-shell particles. Core-shell QDs are also used in magnetic resonance imaging (MRI) for targeted drug delivery [31].

Janus QDs derive their name from the two-faced Roman god Janus, as they exhibit two distinct regions on their surface. These regions can have different surface chemistries, enabling asymmetric functionality. They contain ligands that aid in preserving their shape, size, and performance through a mechanism involving ligand-ligand exchange [32]. Janus QDs possess unique self-assembly properties, allowing them to form complex structures and exhibit tailored interfacial interactions. These properties make them promising candidates for applications such as catalysis, sensing, and biological imaging [33].

Alloyed QDs are composed of different elements within the QD structure. The combination of elements in the QD lattice results in continuous tunability of their optical and electronic properties. Alloyed QDs offer advantages such as broader absorption and emission spectra and the ability to finely tune their bandgaps by controlling the composition, for example, Indium Gallium Arsenide (InGaAs) and $CuInS_2$ [34]. Targeted optical probes made of arginine–glycine–aspartic acid (RGD) peptide-conjugated alloyed QDs are used for *in-vivo* imaging of solid tumors [35].

Doped QDs involve the incorporation of impurities or dopant atoms into the crystal lattice of the QDs. This deliberate introduction of dopants alters the electronic and optical properties of the QDs. Doping can enhance light absorption, improve charge transport, and modify emission characteristics. Doped QDs find applications in optoelectronic devices, including solar cells and transistors. Examples include Mn-doped ZnSe QDs [36] and Eu-doped QDs [7]. The biomedical industry has shown a wide range of uses for doped QDs. It is possible to create Nitrogen-doped carbon QDs with an extremely high quantum yield and add FA to their surface as an anticancer agent [37].

The structural classification of QDs as core-shell, Janus, alloyed, and doped QDs introduces additional dimensions of control over their properties. These structure-based classifications can aid in tailoring the behavior of QDs for specific applications, enhancing their performance, and expanding their potential uses in electronics, photonics, and biomedical applications.

In biological imaging applications (for various cancers such as human ovarian, pancreatic, oral epidermoid, gastric, hepatic, lung, cervical, and ovarian carcinomas), using Gd-doped ZnO, CdTeSe/CdZnS, and CdTe/CdSe/ZnSe has proven to be advantageous [10]. Most common QDs include ZnS QDs, CQDs, Graphene QDs (GQDs), and Magnetically Engineered (ME) QDs.

ZnS QDs show exceptional long-term stability, impressive retention capacity, and enhanced efficacy in pollutant remediation, even at minimal quantities [38]. They are highly efficient, owing to their greater surface area to volume ratio, which improves the absorption of photons at the nano-interfaces [39]. Furthermore, using cost-effective, straightforward, environmentally friendly, and efficient preparation techniques, ZnS QDs can be synthesized efficiently under ambient conditions with high homogeneity, monodispersity, and excellent crystallinity [39]. ZnS has emerged as a widely employed material for fabricating QDs in various photocatalytic applications, including water purification and photo-induced nonlinear optics [40].

CQDs offer good candidates for optronics, catalysis, and sensors because they can operate as electron donors and acceptors, creating electrochemical luminescence and chemiluminescence. They are used as fluorophores (such as organic dyes and fluorescent proteins) due to their high sensitivity, speed, and cost-effectiveness [10,41,42].

Nanohybrid materials incorporating GQDs have garnered substantial attention across diverse research domains, particularly within the biomedical field [43]. The exceptional structural and compositional synergy of graphene makes GQDs an excellent choice for developing versatile immunosensing platforms. Drug delivery systems can be visualized using organic fluorophores and semiconductor QDs to enhance our understanding of cellular uptake [44]. For example, GQD loaded with Gadolinium texaphyrin and lutetium texaphyrin is used to track cell mobility [45]. Electrochemical detection techniques based on GQDs biosensors were widely used in the various nucleic acid assays as they offer a quick, precise, and affordable platform for DNA detection [46].

Magnetically manipulated QDs are versatile multimodal imaging agents combined with contrast agents on a unified nanoparticle (NP) platform. These QDs can be synthesized through various straightforward techniques, such as co-encapsulation, heterocrystalline growth, or assembly of magnetic NPs and pre-formed QDs, doping of QDs with transition metal ions, and conjugation of magnetic chelates onto QDs [47]. Tumour localization using MRI and fluorescence modalities are some of the applications of ME QDs [48].

# Synthesis, Modification, and Characterization of QDS for Biomedical Applications

## *Synthesis of QDs*

This paper primarily concentrates on the applications of QDs, providing only a concise overview of their synthesis methods. For a more comprehensive examination of synthesis techniques, a more detailed review can be found in other sources [49,50]. Two main approaches, bottom-up and top-down, are utilized to create QDs. Top-down processing techniques include ion implantation, molecular beam epitaxy (MBE), e-beam lithography, and X-ray lithography, which involve thinning a bulk semiconductor to form QDs. However, these methods have drawbacks, such as incorporating contaminants and structural flaws [51]. On the other hand, the bottom-up method involves the self-assembly of colloidal QDs in a solution through chemical reduction. Self-assembly methods can be categorized into vapor-phase and wet-chemical processes [52]. A coating based on oxides is necessary for some QDs to reduce toxicity for biological applications since they include harmful ions, including $Cd^{2+}$, $Se^{2+}$, and $Te^{2+}$. The advantages and disadvantages of the abovementioned techniques have been elaborated in Table 2 (Ref. [56-64]). Most QDs used in biomedical applications are constructed with a core made of a particular semiconductor material, such as CdSe, and a shell made of another semiconductor material, such as ZnS. This core-shell design provides several advantages, including improved stability, enhanced optical properties, and reduced toxicity. These techniques synthesize three types of QDs: Type I, Reverse type, and Type II QDs [53–55].

**Table 2. Advantages and disadvantages of different QD synthesis methods.**

| Types of synthesis methods | | Advantages | Disadvantages | References |
|---|---|---|---|---|
| Top-down approach | Ion implantation | - Precise control over dopant concentration<br>- Ensures uniform distribution. | - Limited to certain materials due to high energy requirements<br>- Potential crystal structure damage. | [56,57] |
| | Molecular beam epitaxy | Layer-by-layer growth ensures size and shape control | Limited scalability due to environmental sensitivity and small substrate size constraints. | [58,59] |
| | E-beam lithography | Provides both high precision and compatibility with diverse substrate materials. | Slow large-scale throughput and the risk of charging effects and beam damage. | [60,61] |
| | X-ray lithography | Produces complex patterns with high throughput in large-scale production. | Challenges in achieving sub-10 nm features and ionizing radiation exposure | [62,63] |
| Bottom-up approach | Vapor phase process | - Formation of QD arrays without templates<br>- Ease of self-assembly on substrates without needing additional patterns | - High lattice mismatch causing strain-induced growth of QDs<br>- High concentration of impurities<br>- Fluctuation in size | [64] |
| | Wet chemical process | - Ease of dispersion<br>- Size control<br>- Large-scale production | - High concentration of defects and impurities<br>- High cost | [64] |

***Surface modifications of QDs***

  QDs have shown great promise in various biological applications, but their practical utility is limited by factors such as cadmium release, lead, or arsenic, which are inherent components of many traditional QDs and pose significant toxicity concerns. Additionally, poor solubility hinders their effective utilization in biological systems. Surface modifications have been proposed as a promising solution to address these limitations. One approach is surface ligand exchange, wherein the original toxic ligands on QD surfaces are replaced with biocompatible ligands, effectively reducing the potential release of toxic elements and enhancing biocompatibility. Another strategy involves encapsulating the QDs with biocompatible materials, such as polymer coatings with PEG and poly (D, l, lactide-co-glycolic acid) (PGLA), to provide a protective barrier and improve solubility. Several other innovative methods have also been suggested to address the toxicity and solubility issues in the context of QDs [10,36]. These surface alterations offer potential solutions for enhancing the safety and applicability of QDs in biological systems. Some of these methods have been elaborated in the following paragraphs (Fig. 2, Ref. [74, 75, 76]).

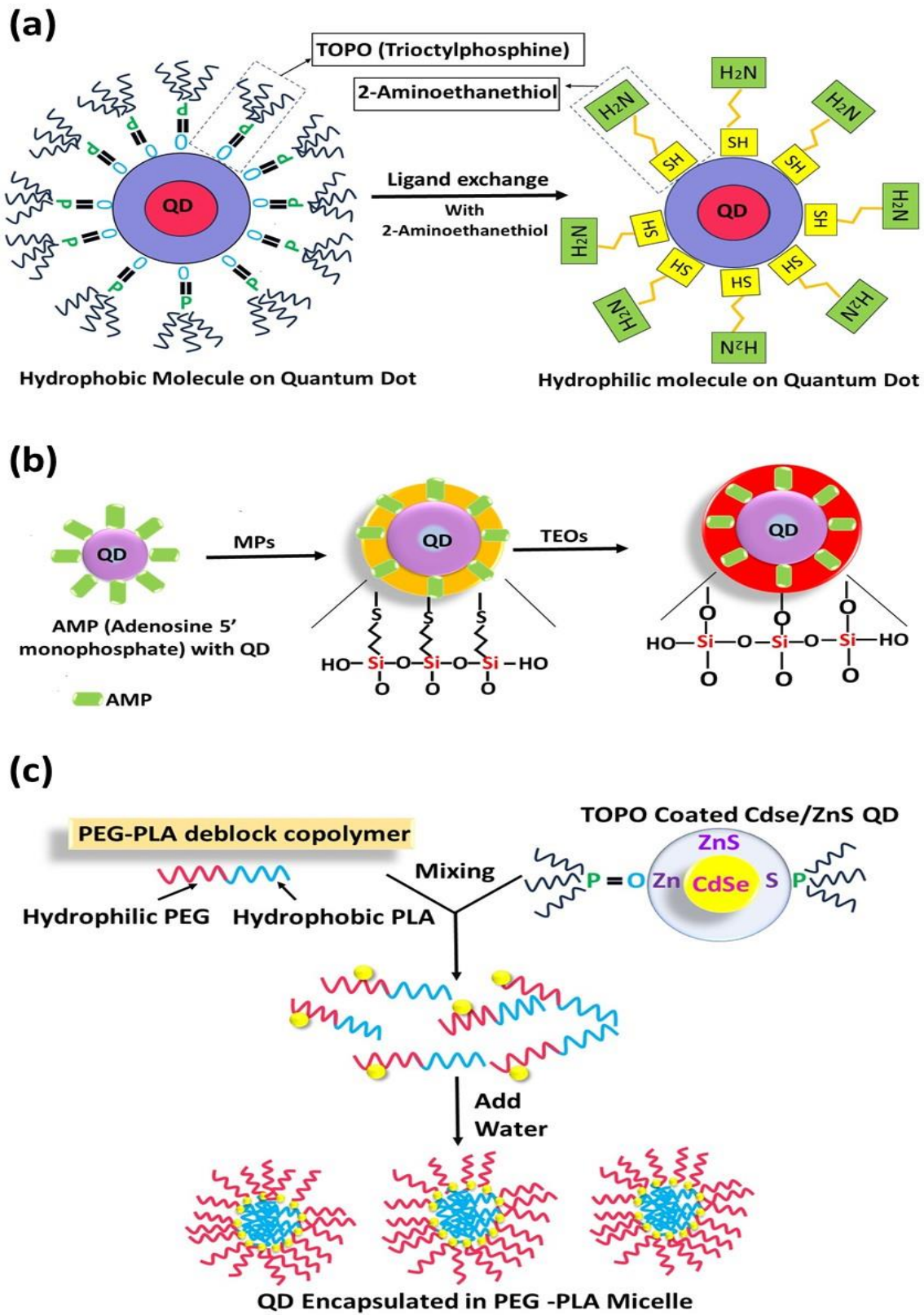

**Fig. 2. Methods of surface modification of Quantum Dots (QDs).** (a) The diagram depicts the surface ligand exchange where a hydrophobic molecule TOPO on QD is being replaced by a hydrophilic molecule 2-Aminoethanethiol, which helps improve the solubility of QD. Adapted from [74] Copyright 2018, Wiley. (b) Silanization is the process by which adenosine 5'-monophosphate (AMP) conjugated Quantum Dots (QDs) are transformed into mercaptopropyltris (methyloxy) silane (MPS), facilitating their solubility in water. Subsequently, these QDs evolve into silica-encapsulated nanospheres through the hydrolysis of tetraethyl orthosilicate (TEOs). Adapted from [75] Copyright 2010, American Scientific Publishers. (c) Encapsulation involves the formation of CdSe/ZnS QD encapsulated using PEG-PLA (poly (ethyl glycol)-b-poly-(D, L-lactide). Adapted from [76]

Copyright 2014, Royal Society of Chemistry. TOPO, Trioctylphosphine oxide; AMP, Aminoethanethiol; MPS, mercaptopropyltris (methyloxy) silane; QD, Quantum dot; PEG-PLA, poly (ethyl glycol)-b-poly-(D, L-lactide); TEOS, tetraethyl orthosilicate; ZnS, Zinc Sulphide; CdSe, Cadmium Selenium QD; Si, Silicon. The diagram was made using Microsoft® PowerPoint® (Version 2310 Build 16.0.16924.20054).

Surface ligand exchange (ap exchange) refers to the process of replacing the native hydrophobic ligands of QDs with hydrophilic ones [65] (Fig. 2a). Commonly used hydrophobic ligands such as trioctylphosphine oxide (TOPO) and trioctylphosphine (TOP), which are frequently present on QDs synthesized using the organometallic method, have been extensively documented in previous studies [66]. The water solubility of QDs can be improved by using various hydrophilic ligands as replacements, such as mercaptocarbonic acids [HS-$(CH_2)$ n-COOH, where n ranges from 1 to 15], thiolated PEG, peptides, oligomeric phosphines, dihydrolipoic acid, dithiothreitol, and 2-aminoethanethiol. However, it is crucial to note that altering the surface ligands can influence the physicochemical properties of QDs and potentially reduce their quantum efficiency [67]. For instance, Silva et al. [68] investigated the effects of different thiol-stabilizing ligands on the characteristics of CdTe QDs synthesized in an aqueous solution.

One widely utilized technique for surface modification is silanization (Fig. 2b), which involves coating QDs with a silica shell to enhance their properties. This method begins by converting the surface ligand, such as trioctylphosphine oxide (TOPO), into mercaptopropyltris (methyloxy) silane (MPS), enabling the QDs to become water-soluble. Additionally, silica nanospheres are generated through the hydrolysis of tetraethyl orthosilicate (TEOS), which are then employed to encapsulate the QDs' surfaces [69,70]. The rate of ligand exchange can greatly influence the fluorescence characteristics of the silanized QDs. Another benefit of this technique is the presence of terminal groups on the silane shell, which allows for additional processing possibilities. However, it is crucial to recognize certain limitations connected with this approach, such as the complexity of the procedure and the necessity for diluted conditions, which may not be feasible for large-scale manufacturing [10].

Another approach for surface modification is surface coatings using polymer encapsulation (Fig. 2c). This method employs polymeric ligands to replace the existing ligands on the surface of QDs [71]. Hydrophobic interactions between the *n*-octyl groups present on QD surfaces (e.g., TOPs or TOPOs) and alkyl chains are utilized in this process. Various materials, including pegylated-phospholipids, Gemini surfactants, cetyltrimethylammonium bromide (CTAB), and amphiphilic oligomers, are used in this process [72]. Low-molecular-weight phospholipid polymers form oil-in-water micelles through hydrophobic interactions with the hydrophobic surface ligands. For instance, Park et al. [73] encapsulated multiple QDs using amphiphilic polyethyleneimine derivatives (amPEIs), demonstrating brighter cell labeling than other QD cellular labeling technique. High molecular weight polymers with hydrophobic and hydrophilic side chains have also been utilized. These amphiphilic polymers offer control over chemical properties, composition, and size tunability [74]. Several polymers have been developed as micelle capsules or ligands to improve the biocompatibility and solubility of semiconductor nanocrystals. However, it is important to note that these methods may result in significantly larger QD diameters than those covered with simple chemical ligands [10].

Surface modifications of QDs offer promising solutions to address the limitations of QDs in biological applications. Methods such as polymer encapsulation, silanization, and surface ligand exchange enhance water solubility, biocompatibility, and stability of QDs. However, careful consideration is needed to optimize these modifications without compromising the optical properties of QDs. Further research in this area is crucial to fine-tune surface modification techniques and unlock the full potential of QDs for safe and efficient use in diverse biological applications.

*Characterization Techniques*

Characterization techniques are crucial in analyzing and understanding QDs and their properties. Various techniques are employed to investigate the structural, optical, and surface characteristics of QDs [77]. One commonly used technique is transmission electron microscopy (TEM), which allows for high-resolution imaging of QDs at the nanoscale. TEM provides valuable information regarding the size, shape, and distribution of QDs, offering insights into their structural properties. Another essential characterization technique is X-ray diffraction (XRD). XRD enables the determination of the crystal structure and lattice parameters of QDs, helping to assess their crystallinity and phase purity. PL spectroscopy is extensively employed to study the optical properties of QDs. PL spectroscopy measures light emission from QDs upon excitation, providing information about their bandgap, emission wavelength, quantum yield, and energy states [78]. UV-visible absorption spectroscopy is used to analyze the absorption properties of QDs in the UV and visible regions. This technique helps determine the absorption spectra, allowing for the calculation of bandgap energies and the evaluation of size-dependent effects. Fourier-transform infrared spectroscopy (FTIR) examines the chemical composition and surface functional groups of QDs. FTIR spectroscopy provides insights into the surface ligands, allowing for the characterization of surface modifications and interactions [79]. Other characterization techniques, such as dynamic light scattering (DLS) and zeta potential measurements, can provide information about the size distribution, stability, and surface charge of QDs [77].

Hazdra *et al.* [80] utilized photo-modulated reflectance spectroscopy, a seldom used experimental technique, to examine deposited QD structures. This approach provides insights into the ground state and multiple higher-order interband optical transitions, allowing for the determination of the band structure, particularly of the wetting (separation) layer. Moreover, it explores a broader range of critical points compared to low-temperature PL [80]. Gu *et al.* [81] used Raman scattering spectroscopy and PL excitation incorporated with a model of PL and longitudinal optical phonon energies to calculate the size and composition of optically active CdZnSe/ ZnBeSe QDs. Field flow fractionation significantly enhances standard characterization methods for water-soluble quantum dots (QDs). Techniques like TEM, atomic force microscopy (AFM), and scanning tunneling microscopy (STM) are commonly employed to monitor the size of epitaxially produced QDs. Magneto-tunneling studies can assess the size of InAs QDs, obtaining results that closely align with AFM and TEM measurements [82]. Only a few techniques have been applied to analyze QD surface chemistry, including Rutherford backscattering, nuclear magnetic resonance (MR), and X-ray photoelectron spectroscopies. Lees *et al.* [83] employed the analytical ultracentrifugation technique to describe the surface chemistry and size distribution of QDs. By detecting changes in nanocrystal sizes that differed by a single lattice plane, it was demonstrated that this approach was incredibly sensitive to nanocrystal size [84]. The combination of these characterization techniques enables a comprehensive understanding of the structural, optical, and surface properties of QDs, facilitating their optimization and application in various fields.

## Biomedical *In-Vivo* Imaging Using QDS

QDs have emerged as valuable biomolecular and cellular imaging tools, offering unique optical properties that set them apart from traditional organic dyes and fluorescent proteins. These tiny light-emitting particles have the potential to revolutionize biomedical imaging by providing enhanced visualization capabilities. One of the remarkable advantages of QDs is their exceptional brightness and tunability across different colors [85]. Unlike other fluorescent probes, QDs exhibit prolonged photostability, retaining their brightness over extended periods. This characteristic allows the development of highly sensitive tools for investigating various biological processes within living organisms [86]. Traditional fluorescent labels, such as organic dyes and fluorescent proteins, have long been used for studying living things. However, they suffer from brightness loss and color variability when used over extended periods or with multiple colors. Alternate fluorescent labels have been developed to overcome these limitations [87]. QDs, recently developed as fluorescent NPs, exhibit strong fluorescence, resistance to photobleaching, and the ability to generate multiple colors [88].

### *Fluorescent QD Probes*

In stimulated emission depletion (STED) nanoscopy, different NPs and organic dyes are used as fluorescent probes. However, these probes have problems like being toxic, quickly losing their brightness, and not dissolving in water. To solve these issues, researchers have developed a new type of fluorescent material called fluorescence carbon dots (FNCDs). These FNCDs have a high brightness level, are non-toxic, do not lose their brightness quickly, and can dissolve in water. They have proven effective in visualizing specific components of living cells, such as the nucleolus, and can also reduce unwanted background signals, enhancing image clarity [89]. QDs can bind to various targets like antibodies and peptides. They promise enhanced sensitivity and specificity in cancer detection compared to conventional methods. However, toxicity and nonspecific uptake still impede their widespread adoption [90].

QDs-based signal probes have been used in many biotechnological applications [90]. Fluorescent semiconductor QDs are utilized as markers for single-stranded DNA sequences that complement plasmid clusters. These bioconjugate probes demonstrate excellent dispersibility, bioactivity, and specificity for hybridization. By employing the fluorescence in-situ hybridization (FISH) technique, successful localization of plasmid pUC18 clusters within Escherichia coli at the subcellular level has been achieved [91]. They have been effectively employed in immunofluorescence labeling of fixed cells and tissues, staining membrane proteins, microtubules, nuclear antigens, and FISH of chromosomes [90].

QDs offer advantageous optical properties for *in-vivo* imaging of live cells or tissue in animals. Overcoming challenges like immune response, toxicity, and biomolecule attachment is crucial [92]. Studies have shown the potential of QDs in *in-vivo* imaging, with persistence observed in mice for up to seven days and internalization by various cell types, including neurons. This approach aids in understanding neuronal response and QD toxicity. The study employed transgenic mice monitored for an astrocyte-responsive luciferase reporter and QD conjugate injections, revealing increased bioluminescence as a response [89]. *In-vivo* targeting studies using QD probes have demonstrated delivery to tumor sites through both passive and active targeting mechanisms in nude mice with human prostate cancer. Multifunctional QD probes injected systemically and subcutaneously injected QD-tagged cancer cells enabled multicolor fluorescence imaging of as few as 10–100 cancer cells *in vivo* [93].

An example of *in-vivo* optical imaging is the subcutaneous introduction of CQDs into a nude mouse is shown in Fig. 3 (Ref. [94]). The mouse was imaged under wavelengths 455, 523, 595, 605, 635, 661, and 704 nm (Fig. 3a). Two types of fluorescence were observed, green and red. The non-cancerous tissue and C-dots in mice showed green (autofluorescence) and red fluorescence, respectively. The clarity and contrast between the two were best at 704 nm. Fig. 3b represents a plotted graph with a fluorescent intensity of C-dots and autofluorescence on the Y-axis and the wavelengths used for imaging at the X-axis. C-dots showed highest intensity between 750 and 800 nm, indicating that for *in-vivo* imaging of C-dots in nude mice, a wavelength of 750–800 nm is the most suitable as it provides the best clarity and helps in the precise localization of C-dots [94].

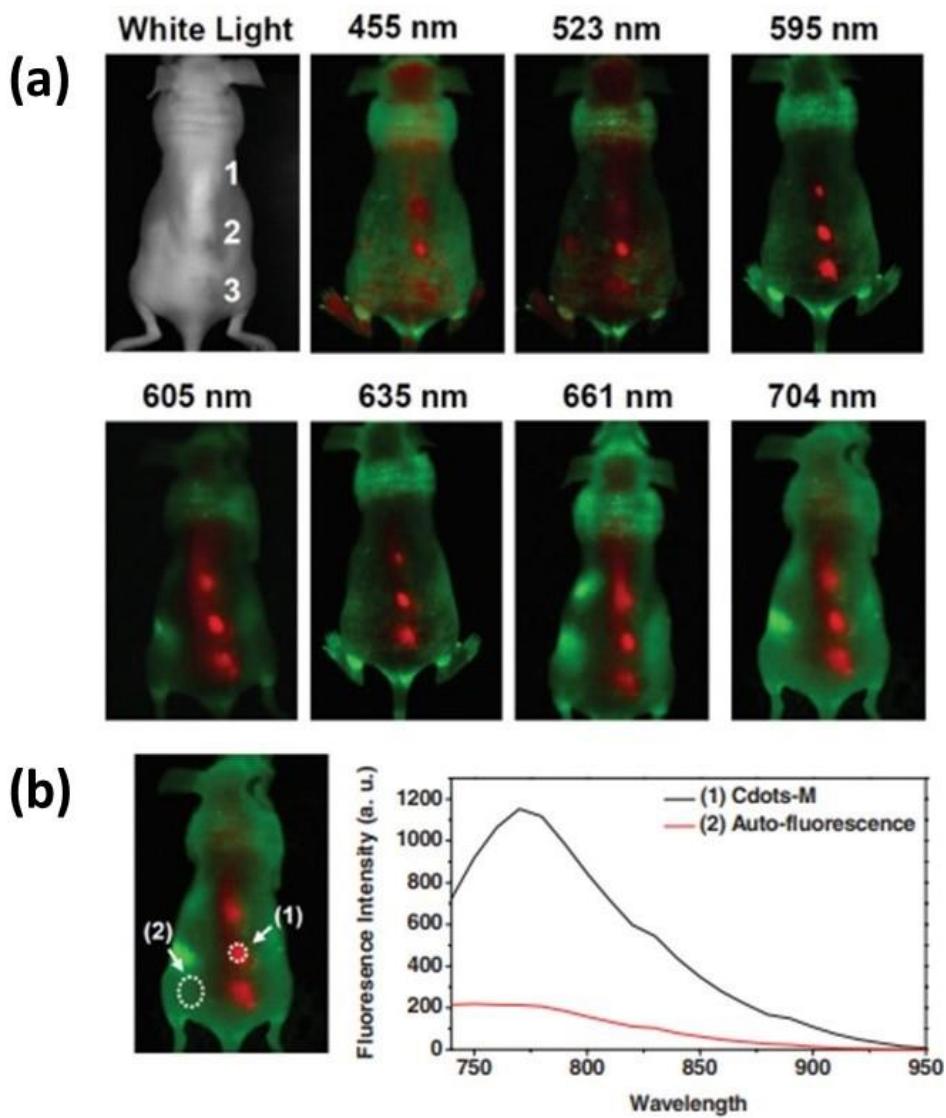

**Fig. 3.** (a) *In-vivo* fluorescence imaging results at different wavelengths of 455, 523, 595, 605, 635, 661, and 704 nm when CQDs are injected in mice. Red shows the fluorescence from CQDs, and green shows tissue autofluorescence. (b) The CQDs can be easily differentiated from tissue autofluorescence at 704 nm excitation. Reproduced with permission [94] Copyright 2012, Wiley. CQDs, Carbon QDs.

In summary, the unique optical properties of QDs, such as their brightness, photostability, and tunability, make them valuable tools for *in-vivo* imaging. However, further research is required to address toxicity, specificity, and efficient targeting challenges, paving the way for their broader application in biomedical imaging.

*Multimodal Imaging*

Multimodal imaging probes have gained significant interest in biomedical research by combining the unique properties of light-emitting semiconductor QDs with MRI in a single NP [48]. These hybrid materials serve a dual purpose, allowing for fluorescence imaging and MRI, providing complementary information for comprehensive imaging [95].

Various strategies have been explored to develop contrast agents integrating QDs with MRI capabilities to achieve multimodal imaging. One approach involves incorporating a layer of magnetic material onto the surface of QDs, enabling fluorescence emission and MR contrast enhancement [96]. Another strategy involves introducing paramagnetic ions to the QD structure, which enhances the contrast for MRI while retaining the fluorescent properties of QDs [96]. Additionally, using silica or polymer NPs as a matrix to encapsulate QDs and paramagnetic ions has been explored to create multimodal contrast agents. Chelating molecules can also be incorporated to bind paramagnetic ions to QDs, further enhancing multimodal imaging capabilities [96].

The development of multimodal imaging probes offers several advantages in bioimaging. Structural and functional information can be obtained simultaneously by combining fluorescence imaging with MRI, leading to a more comprehensive understanding of biological processes [47]. These probes can provide high-resolution fluorescence images to visualize specific cellular components, while the MRI component enables detailed anatomical and physiological imaging of tissues [96].

An example of multimodal imaging involves three modalities: intravital microscopy (IVM), MRI, and whole-body fluorescence. Images of mice are captured after injecting with RGD-pQDS (RGD peptide-paramagnetic QDs). IVM identifies tumor and angiogenesis at the microscopic level of superficial blood vessels (Fig. 4a, Ref. [97]). When the same mice were imaged using MRI, the tumor's maximum angiogenic activity, highest resolution, and three-dimensional information were visible (Fig. 4b). Whole-body fluorescence, which is a very sensitive method allows for quicker screening of NP targets in the entire body of mice (Fig. 4c) [97].

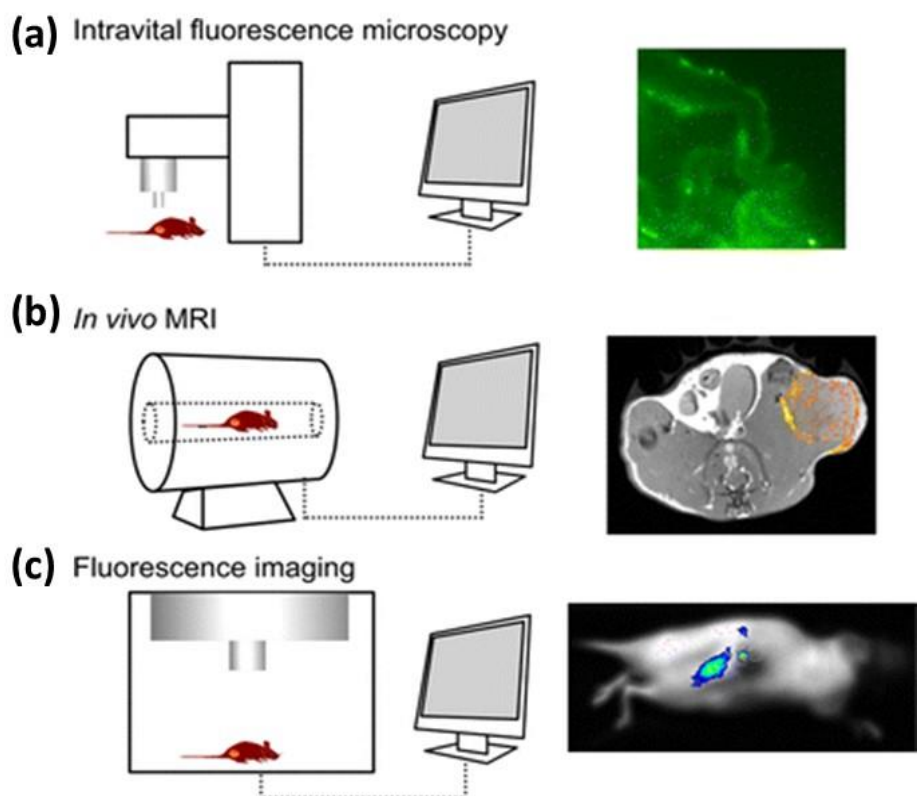

**Fig. 4. Different methods involved in multimodal imaging: their instrumentation and output image.** (a) Intravital fluorescence microscopy (IVM) is used to image QDs in superficial blood vessels. (b) Magnetic Resonance imaging (MRI) gives high-resolution three-dimensional information about the location of QDs. (c) Whole-body fluorescence imaging helps screen QDs in the entire mice body. Reproduced with permission [97] Copyright 2010, Springer. MRI, Magnetic Resonance Imaging; IVM, Intravital fluorescence microscopy.

CQDs, in combination with nitrogen and lanthanides, have been successfully synthesized and investigated as multimodal contrast agents for imaging applications. These NPs exhibit exceptional light-emitting properties and demonstrate strong signals in MRI and computed tomography (CT) scans. Importantly, they have demonstrated excellent biocompatibility, as evidenced by their non-cytotoxic nature when exposed to different cell lines for extended periods (24 and 72 h), making them highly suitable for multimodal imaging applications [98].

Integrating multiple imaging functions into a small and biocompatible nano platform has posed a significant challenge in developing multimodal imaging probes. However, novel CQDs doped with gadolinium and ytterbium (Gd/Yb@CDs) were successfully synthesized using a simple hydrothermal process. These Gd/Yb@CDs have demonstrated exceptional MRI, X-ray CT, and fluorescent imaging (FI) capabilities. Notably, they exhibit emission characteristics dependent on excitation, remarkable photostability, high longitudinal relaxivity, and excellent X-ray absorption performance [99].

Core/shell $CdSe/Zn_{1-x}Mn_xS$ NPs have been developed as effective optical and MRI agents. These NPs exhibit high luminescence and quantum yield in organic solvents and a practical low yield in water. Moreover, they possess high relaxivity, making them suitable for MR imaging. Successful optical and MR imaging studies have been conducted using these particles in aqueous solutions and cell cultures, highlighting their potential for medical imaging applications at low concentrations [100].

Microvesicles (MVs), tiny bubbles released by cells, hold great potential for delivering treatments and carrying important substances between cells. $Ag_2Se$@Mn QDs are tiny particles developed to label and track MVs effectively. These QDs enable high-resolution dual-mode tracking using fluorescence and magnetic imaging *in vivo*, offering valuable insights into MV behavior and distribution [101].

Furthermore, a straightforward approach for fabricating Mn: ZnSe/ZnS/ZnMnS sandwiched QDs (SQDs) has been discovered, which constitute layers containing paramagnetic $Mn^{2+}$ ions, enabling Mn phosphorescence and enhancing MRI signals. This innovative approach holds promise for advanced multimodal imaging applications, leveraging the unique properties of SQDs [102]. Developing these innovative nanoplatforms and QD configurations highlights the continuous progress in creating highly functional multimodal imaging probes. These advancements offer exciting prospects for improving imaging capabilities, enhancing diagnostic accuracy, and facilitating targeted therapies in biomedical research and clinical settings.

## QDS as Drug Delivery Systems

Nanoparticle-mediated drug delivery (NDD) demonstrates great potential in enhancing drug efficacy and facilitating the development of innovative therapies. Numerous studies have highlighted the benefits of NDD systems, including targeted delivery, controlled drug release, extended circulation times, improved bioavailability, and reduced drug toxicity. However, a model system that can accurately evaluate all stages of NDD while possessing high sensitivity, resolution, and cost-effectiveness is urgently required. QDs serve as an excellent model system for monitoring nanocarriers' clearance, drug release, degradation, and distribution [103]. They effectively maintain the characteristics of the original NDD carriers. These particles can be utilized to gain insights into creating improved and safer drugs for both humans and animals. By emitting light, these particles help visualize their movement within the body, surpassing the limitations associated with harmful chemicals [104].

### QDs for Theranostics

QDs exhibit remarkable optical characteristics, encompassing elevated quantum yields, expansive absorption spectra, and finely adjustable emission bands, making them prime candidates for bioimaging and diagnostics [105,106]. Harnessing the expansive surface-to-volume ratio of QDs, a sophisticated multifunctional nanoplatform has been crafted. QDs are imaging agents and nanoscaffolds for diverse therapeutic and theranostic applications [107]. In a recent investigation, a cancer theranostics platform showed biocompatible $Fe_3O_4$-ZnO core-shell MQDs to visualize and treat tumors simultaneously. This platform showcased distinctive traits, such as the generation of singlet oxygen and superparamagnetism [108]. The inventive strategy seamlessly integrates radiation and photodynamic therapy with MR imaging, harboring transformative potential for cancer theranostics. An alternative study introduced CdTe QDs theranostic nanocapsules, featuring a positively charged oily core laden with rapamycin and celecoxib, enveloped in a chondroitin sulfate coat [109]. An external layer of cationic gelatin-coupled QDs was introduced to avert nonspecific uptake, ensuring precision in tumor targeting. Moreover, an unconventional approach employing the iron-binding cationic protein lactoferrin induced an ON-OFF effect—initially dampening QD fluorescence and restoring it post-bond cleavage within tumor cells [110]. This diverse approach sheds light on theranostics, demonstrating the versatile role of QDs beyond conventional applications.

GQDs exhibit versatility and find applications in various fields, including drug delivery, sensors, bio-imaging, and environmental protection. GQDs offer several advantages compared to other NPs, such as low toxicity, a large surface area, and adjustable PL [111]. They hold promise in nanomedicine as drug carriers due to their enhanced water solubility and reduced cytotoxicity compared to graphene. They also provide more bonding sites for chemotherapeutic conjugation and demonstrate improved cellular uptake [112]. Carbon allotropes, like graphene, can be used for drug delivery systems and imaging probes. GQDs possess unique properties, such as stable fluorescence and multimodal conjugation, which make them highly effective and less toxic as delivery systems. Recent studies have shown that GQDs can enhance the efficacy and cytotoxicity of anticancer drugs, positioning them as a promising tool for cancer therapy [113]. GQDs, with their planar structure, facilitate efficient drug loading and delivery. They have even demonstrated the ability to cross the blood-brain barrier without surface functionalization, potentially aiding in the prevention of α-synucleinopathy in Parkinson's disease. The size and charge of NPs are crucial factors influencing their ability to permeate the blood-brain barrier, with smaller NPs exhibiting better permeability and diffusion in the brain [112].

N-doped carbon QDs (N-CQDs) also hold potential as drug carriers and imaging agents. In a study, N-CQDs were synthesized hydrothermal, conjugated with quinic acid, and loaded with gemcitabine, an anticancer drug. These NPs were evaluated for their luminescent properties, cell viability, and imaging capabilities in breast cancer cells in mice. The quinic acid-conjugated N-CQDs exhibited excellent luminescence and high tumor accumulation, suggesting their potential as multifunctional theranostic agents [114]. CQDs have emerged as a promising nanomaterial for cancer imaging and drug delivery. They possess desirable characteristics such as fluorescence emission, photostability, and biocompatibility. Recent studies have explored various CQD-based anticancer drug delivery strategies, including active and passive targeting, multifunctional targeting, and targeting the tumor microenvironment [115]. Moreover, developing advanced CQDs for *in-vivo* imaging and biodistribution in cancer model NPs holds potential for future advancements in cancer treatment [41]. NPs have been employed to co-deliver drugs and imaging agents in cancer therapy. In one study, the co-loading of Paclitaxel and CdTe/CdS/ZnS QDs into nanostructured lipid carriers inhibited tumor growth. Another study utilized hybrid silica nanocapsules loaded with ZnSe: Mn/ZnS core-shell and Paclitaxel for chemotherapy and fluorescence imaging, improving Paclitaxel solubility and sustained drug release [116]. pH-responsive ZnO QDs loaded with Doxorubicin enabled drug release in acidic intracellular conditions. Additionally, Quercetin-loaded CdSe/ZnS QDs exhibited enhanced efficacy against drug-resistant bacteria and cancer cells compared to Quercetin or CdSe NPs alone [1]. Carbon nanomaterials, particularly carbon nanotubes (CNTs), have gained popularity as carriers for drug delivery. CNTs are considered safe due to their inert chemical composition. The unique spectroscopic properties of CNTs, such as Raman and PL, provide valuable tools for tracking, detection, and imaging to study the behavior of nanocarriers and assess the effectiveness of drug delivery in living organisms [117].

*Targeted Drug Delivery Using QDs*

Targeted drug delivery and early detection of cancer cells pose significant challenges in cancer treatment. Current approaches involve drug delivery systems that specifically target tumors, consisting of a component that recognizes the tumor and vesicles loaded with drugs. Detection methods typically involve invasive techniques like tissue biopsy or advanced diagnostic tools such as MRI. However, no known system currently can simultaneously target drug delivery and imaging the delivery process. The integration of biomaterials and semiconductor nanocrystal QDs shows promise in cancer therapy. This approach utilizes biodegradable chitosan for tumor-targeted drug delivery and QDs for non-invasive imaging [48,118].

Using natural sugars in the synthesis of QDs offers a promising approach to enhance the safety and effectiveness of QDs for imaging applications. One method involves mixing Mn: ZnS QDs with chitosan, which reduces their potential harm and allows FA attachment. The resulting nano-delivery system, FACS-Mn-ZnS, exhibited no toxicity against breast cancer and normal breast cells [119]. That highlights the significant potential of natural sugars in improving the safety and efficacy of QDs for imaging purposes. Furthermore, the development of Zein-ZnS QD nanohybrids shows promise for drug delivery, particularly for the targeted release of drugs like 5-fluorouracil [119]. In bioimaging, protein-QD nanohybrids play a crucial role and demonstrate stability in biological fluids. The coupling of QDs with proteins can be achieved through chemical reactions or physical entrapment. For instance, GQDs have been successfully linked with human serum albumin NPs loaded with gemcitabine for effective imaging [119].

Researchers continuously seek better ways to deliver anticancer drugs to specific sites and explore new nanomaterials for improved drug delivery. They have identified new drug delivery-release modes, such as enhanced permeability and retention (EPR)-pH delivery-release mode (Fig. 5a), ligand-pH delivery-release mode (Fig. 5c), EPR-Photothermal Delivery-Release Mode (Fig. 5b), and Core/Shell-photothermal/magnetic thermal delivery-release mode (Fig. 5d). These modes are commonly used for cancer treatment, while others are utilized for non-tumor diseases [112].

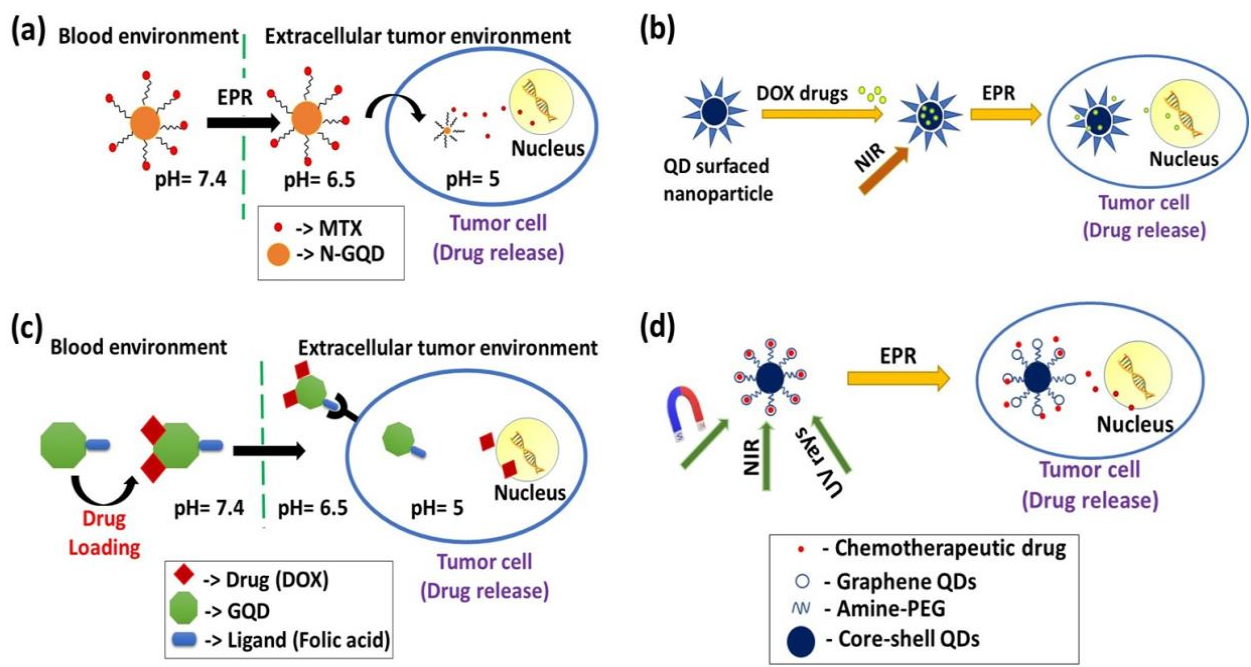

**Fig. 5. Modes of drug delivery by Quantum Dots (QDs) to treat cancer.** (a) EPR pH Mode, which involves pH-responsive drug delivery near the acidic extracellular tumor environment and exploits the Enhanced Permeability and Retention effect of tumor tissue. (b) EPR-Photothermal Delivery-Release Mode uses near-infrared loading of the drug onto the QD. (c) Ligand-pH delivery-release mode works by a ligand attached to the QD, which selectively binds to tumor cells and releases the drug. (d) Core/Shell-photothermal/magnetic thermal delivery-release mode wherein the drug is loaded by techniques such as magnetic, photothermal, and radiation. EPR, Enhanced Permeability and Retention effect; MTX, Methotrexate; N-GQD, Nitrogen-doped Graphene Quantum dot; DOX, Doxorubicin; GQD, Graphene Quantum Dot; NIR, Near Infrared Radiation; UV, Ultraviolet; Amine-PEG, Amine group conjugated with Polyethylene Glycol; QD, Quantum Dot. The diagram was made using Microsoft® PowerPoint® (Version 2310 Build 16.0.16924.20054), and the DNA structure was taken from stock images available on PowerPoint®.

A pH-responsive drug delivery system utilizes luminescent aminated ZnO QDs as nanocarriers. The ZnO QDs are stabilized with PEG and targeted to cancer cells using hyaluronic acid (HA) [120]. Gu *et al.* [121] conducted a study on small-cell lung cancer and developed a targeted drug delivery (TDD) system using Ag-In-Zn-S QDs modified with various compounds, including FA. The TDD system specifically targets FA receptors on cancer cells and successfully delivers doxorubicin to adenocarcinomic human alveolar basal epithelial cells, demonstrating its potential as a novel treatment option for lung cancer. However, further studies must evaluate its safety and effectiveness *in vivo* [122]. Fullerene, particularly a modified form called C60-PEI-doxorubicin (DOX), shows promise in drug delivery and photodynamic therapy by combining chemotherapy and photodynamic therapy in one treatment approach [123]. Tumors often exhibit low pH levels, and the use of a traceable and pH-responsive drug delivery system called $MoS_2$-PEG-DOX can be used, which is based on $MoS_2$ QDs covalently linked with diamine-terminated oligomeric PEG and loaded with the fluorescent anticancer drug doxorubicin (DOX) [124].

QDs have a lot of potential in diagnosing and treating human diseases. However, most NPs face challenges related to their size and charge, impeding their efficient elimination from the body. This limitation presents a potential risk of increased toxicity. There are three crucial factors to consider when evaluating whether an NP has potential for use in medical treatments: (a) Size: which should be less than 5.5 nm; (b) Non-toxicity, to avoid any harmful effects to the body and (c) Biodegradability [125].

QDs interact with plasma proteins and blood cells when they enter the bloodstream, affecting their distribution and elimination. The cellular uptake mechanisms for QDs might include diffusion and transmembrane processes (Fig. 6). Inhalation of these QDs leads to their accumulation in the lungs and can cause inflammation [126]. Animal studies have shown their presence in organs like the liver, heart, kidney, spleen, and brain [127]. After absorption by the gastrointestinal tract through oral administration, QDs are transported to the liver via the portal vein and may be actively eliminated. The QDs conjugated with drugs undergo metabolism similar to conventionally formulated drugs. The drugs get intracellularly absorbed, whereas the QDs enter the systemic circulation and are excreted through renal and fecal clearance [126]. Studies on mice intravenously injected with QDs show the presence of QDs in the lymph nodes and bone marrow for an extended period, indicating a potentially long half-life [128,129]. While their excretion through sweat glands and breast milk may also be possible, data confirming these pathways is unavailable. Long-time exposure to accumulating QDs may pose risks, although low concentrations or single exposures are less harmful. Advanced research is required to understand their administration, accumulation, and excretion routes [126].

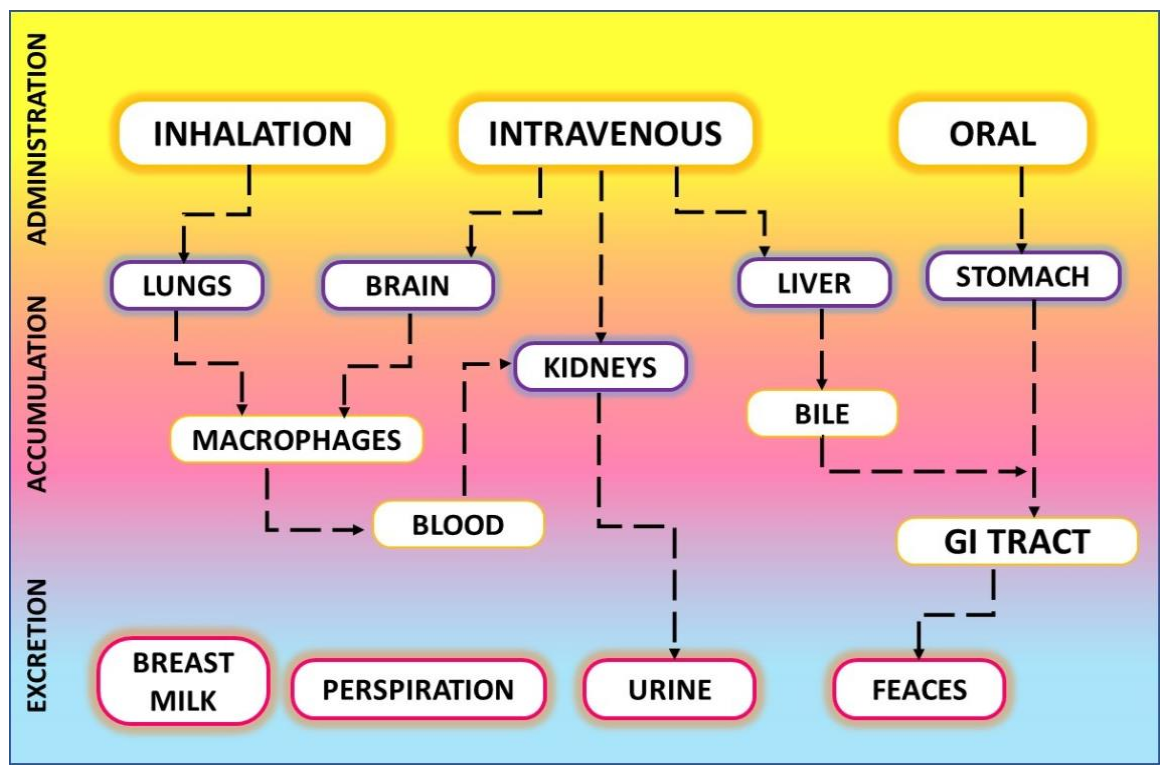

**Fig. 6. Different modes of excretion of Quantum Dots (QDs).** Renal and fecal methods are the most common ways of excretion. Perspiration and breast milk can also aid in the removal of QDs. This figure illustrates the pathways of QDs following inhalation, intravenous injection, and oral administration. Inhalation leads to accumulation in the lungs. Intravenous injection primarily targets the liver, kidneys, and brain. Oral administration results in gastrointestinal accumulation. The liver plays a vital role in the metabolism and removal of QDs. Excretion pathways include renal clearance through urine, biliary excretion through feces, and potential excretion through breast milk and sweat glands. The diagram was made using Microsoft® PowerPoint® (Version 2310 Build 16.0.16924.20054).

## QDS for Biosensing and Diagnostics

### QDs in biosensing

QDs possess unique properties that make them ideal for biosensing and bioimaging applications, including narrow emission bands and broad absorption, easily tunable optical properties, long fluorescent lifetime, resistance to photobleaching, and high specific surface area [130]. GQDs are particularly sensitive, stable, and selective, making them useful as PL probes for detecting inorganic and organic molecules and biomacromolecules in biological samples. GQDs offer advantages over traditional dyes in terms of low toxicity and biocompatibility, and they have been employed to identify cancer cell lines [131,132]. QDs play a significant role in biosensing, including immunoassays, nucleic acid detection, and Förster resonance energy transfer (FRET). In immunoassays, QDs serve as probes for sensing chemical and biological analytes. For example, CQDs conjugated with antibodies specific to myocardial troponin I have been used to detect acute myocardial infarction (AMI) [133]. Luminescent QD beads have been utilized in the detection of Hepatitis B surface antigen HBsAg in human serum and the detection of aflatoxin B1 in maize using competitive immunochromatographic assays [134].

Studying RNA and DNA in cells is crucial for understanding cell behavior and diagnosing diseases. QDs can detect pathogenic gene fragments of different microorganisms in biological samples, aiding in disease diagnosis [135]. Cationic QD probes are useful for detecting and distinguishing nucleic acids (RNA and DNA) in cells. These probes can easily penetrate various biological barriers compared to dyes, which are limited to crossing cellular-level barriers. CQDs can bind to single-stranded RNA (ssRNA) and double-stranded DNA (dsDNA), producing different fluorescence spectra for each [136]. DNA-bound QD probes have been used not only for detecting miRNAs through hybridization techniques but also for assessing the levels of certain chemicals, such as dopamine, in humans [133]. FRET is a technique that involves the transfer of energy between two fluorescent molecules (donor and acceptor) when they come into proximity, as shown in Fig. 7. QDs can be used in FRET systems to detect the presence of organic phosphorothioate pesticides, nicotine insecticides, strobilurin fungicides, and quinolone antibiotics in fruits, vegetables, and honey. Oligonucleoside aptamers (APTs) can bind to QDs and be utilized in FRET-based detection of pesticides and veterinary drugs like chloramphenicol. Here, QDs act as the donor or acceptor molecule [137–139].

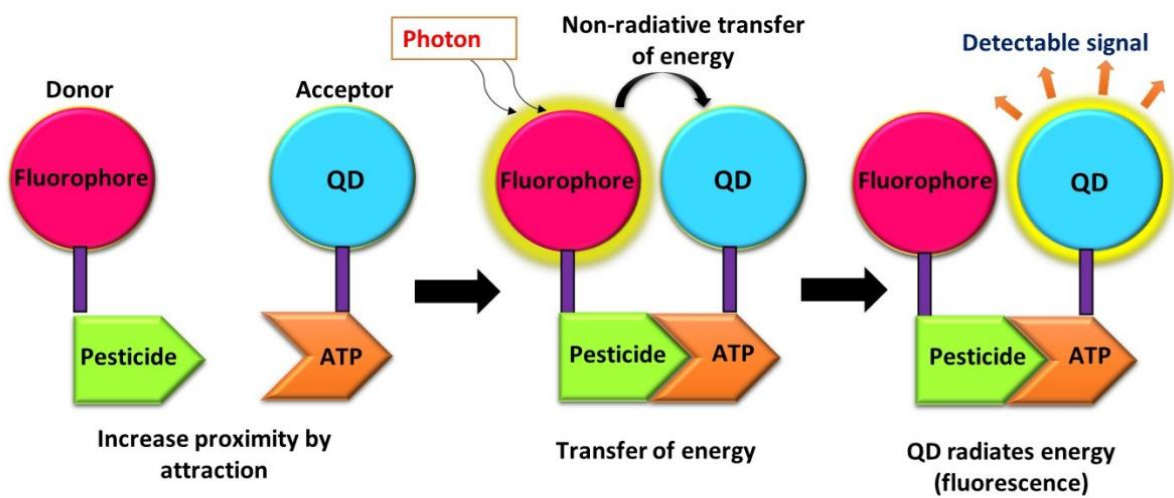

**Fig. 7. Foster Resonance Energy Transfer (FRET) imaging is used to detect the presence of target compounds by a fluorophore that transfers energy to QDs whose fluorescence can be measured and observed.** ATP, Adenosine Tri Phosphate; QD, Quantum Dots. The diagram was made using Microsoft® PowerPoint® (Version 2310 Build 16.0.16924.20054).

*QDs for Disease Diagnosis*

QDs offer numerous advantages for biomedical imaging applications, including detecting pathogens, toxins, and cancer cells and assessing drug efficiency and delivery for various diseases affecting organs like the liver, kidney, and brain [140]. *In-vitro* imaging using QDs allows for detecting tumor cells, stem cells, and other targeted cells. QDs can easily penetrate stem cells without affecting their viability, making them valuable for studying these cells. Additionally, QDs can image human breast cancer cell lines, macrophages, hepatocellular cells, and intracellular components such as microtubules and organelles [130,141]. QDs coupled with specific antibodies can bind to overexpressed receptors, such as the Her2 receptor in breast cancer cells or epidermal growth factor (EGF) receptors in ovarian cancer, enabling the detection of these cancer types [130]. *In-vivo* imaging with QDs offers a cost-effective alternative to traditional methods like MRI, PET, and ultrasound for detecting cells and analytes in animal models such as mice. Different types of QDs, such as green carboxylated GQDs for superficial tissue imaging and red-emitting GQDs for imaging HeLa cells, can be used depending on the desired application [142]. N-GQDs with near-infrared absorption properties have high penetrating power and are suitable for deep-tissue cell detection [132]. QDs have been used for real-time imaging of embryo development in Xenopus frogs and for visualizing vasculature, including capillary networks and lymph nodes, in organisms [130]. They can also be used for tumor cell imaging [143].

QDs can also be used to detect bacterial infections without needing antibodies. QDs can label bacterial DNA, effectively acting as fluorescent barcodes that enable the identification of distinct bacterial strains [144]. Enzyme detection plays a crucial role in disease diagnosis; however, individually detecting enzymes through enzyme-substrate reactions can be costly and time-consuming. QDs can facilitate multi-enzyme detection by acting as fluorescent markers, providing high sensitivity and multiplexing capabilities, allowing multiple enzymes to be monitored simultaneously using the same polypeptide substrate [145]. Mitochondrial dysfunction is associated with various diseases, and QDs, particularly single-layered GQDs, can serve as fluorescent probes to attach to the mitochondrial surface and detect changes in mitochondrial membrane potential [146].

# Toxicity and Safety Considerations of QDs in Biomedical Applications

## Toxicity Mechanism and Toxicity Evaluation of QDs

While QDs have shown promising results in biomedical imaging, it is vital to consider their potential toxicity. QDs can cross biological barriers or enter the body through various routes, such as inhalation, ingestion, or intravenous administration, and reach organs where they may exert toxic effects [147]. *In-vitro* studies provide an initial understanding of potential toxicity, but *in-vivo* studies are necessary to comprehensively assess the adsorption, distribution, metabolism, and excretion (ADME) of QDs [148]. QDs can vary depending on their size, with larger QDs being more toxic because they cannot be easily excreted and can accumulate in the system for extended periods [147]. Fig. 8 shows the effect of the toxicity of QDs on different organs.

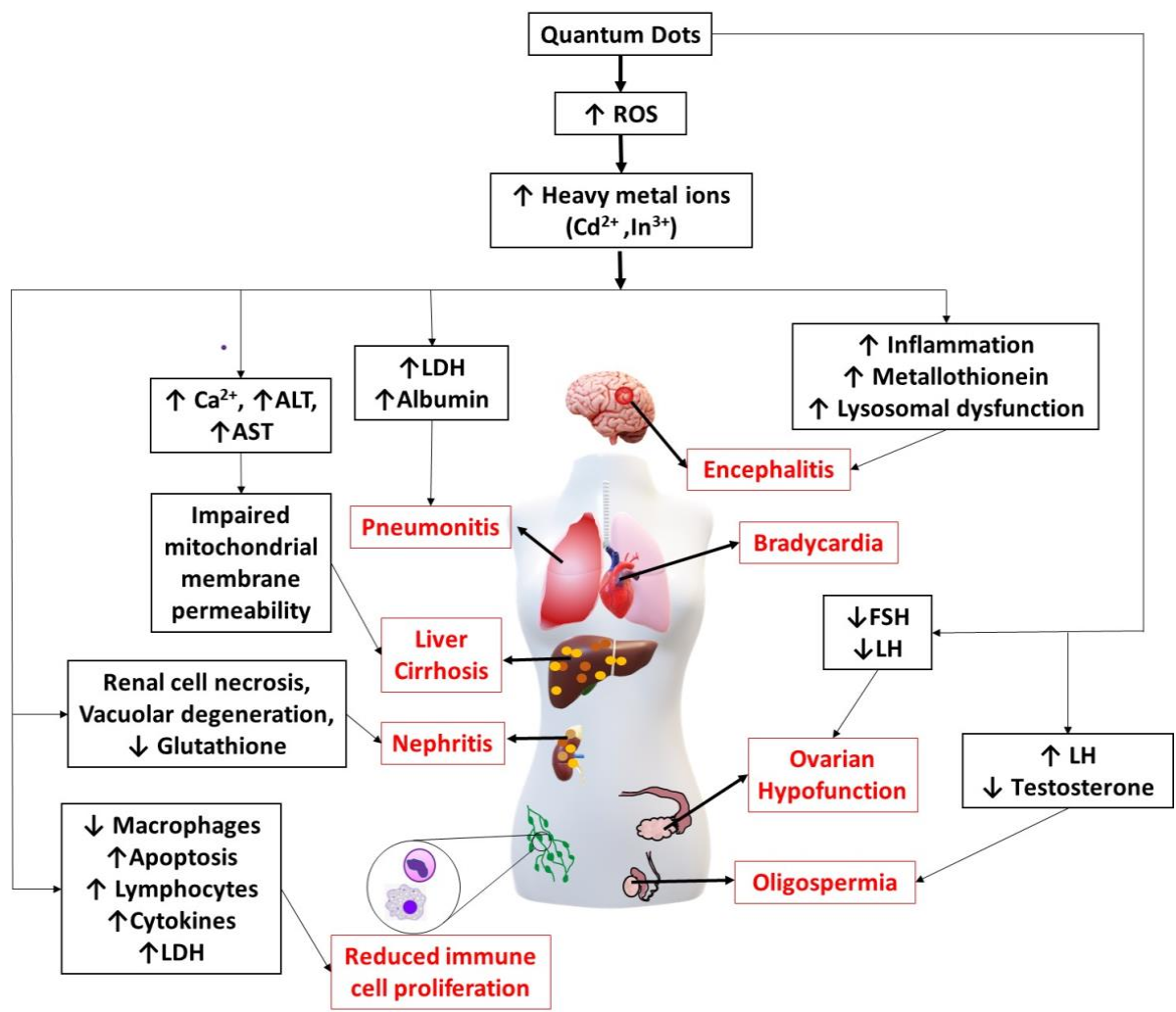

**Fig. 8**. **Diagram illustrating the multifaceted mechanisms involved in the toxicity of Quantum Dots (QDs) in various organs and associated disease pathogenesis.** QDs travel through the blood and accumulate in several organs like the liver and kidneys, leading to inflammatory diseases like cirrhosis and nephritis, respectively. When inhaled, QDs can cause pneumonitis in the lungs. They can cross the blood-brain barrier and cause encephalitis. Even reproductive organs are affected, leading to low fertility. Incorporation of QDs in the lymph nodes can cause reduced proliferation of immune cells like macrophages and lymphocytes. ROS, Reactive oxygen species; LDH, Lactate dehydrogenase; ALT, Alanine aminotransferase; AST, Aspartate aminotransferase; FSH, Follicle-stimulating hormone; LH, luteinizing hormone. The diagram was made using Microsoft® PowerPoint® (Version 2310 Build 16.0.16924.20054). Individual images of organs were taken from stock images available on PowerPoint®.

QDs conjugated with heavy metals like $Cd^{2+}$ can be toxic to the liver. QDs can induce the formation of reactive oxygen species (ROS) and cause lysozymes breakdown, releasing Cd2+ ions that can damage intracellular proteins and DNA structure [149]. QDs can also increase intracellular $Ca^{2+}$ levels, impairing mitochondrial membrane permeability and apoptosis. Elevated levels of liver enzymes such as alanine aminotransferase (ALT) and aspartate aminotransferase (AST) indicate liver damage. CdSe QDs of size 4 nm, at a dosage level of 5–10 nmol per mouse, were observed to be toxic to the liver [150,151]. QDs can accumulate in the lungs through pulmonary inhalation and cause serious complications. CdSe/ZnS QDs, for example, can increase oxidative stress, lactate dehydrogenase, and albumin levels in the lungs, damaging lung tissue. Mice injected with 10nmol kg$^{-1}$ of these QDs resulted in a 100 percent mortality rate. Dissolution of QDs can release cadmium ions that affect cell membranes by disrupting potassium channels. Certain types of QDs, such as PL carboxylated GQDs and silicon-based QDs, are known to be carcinogenic to respiratory cells [147,152]. QDs can be metabolized by the kidneys, releasing heavy metals such as Cadmium and Indium, which can accumulate and cause renal tubular epithelial cell necrosis, vacuolar degeneration, and deformation of glomeruli, nuclei, and tubules. CdSe QDs and InGaP QDs, at a concentration of 10 and 100 nmol, respectively, can induce metal-induced oxidative stress, autophagy, and a decrease in glutathione content in renal cells [153,154]. QDs can easily cross the blood-brain barrier and reach neurons, posing a significant health risk. Accumulation of QDs in neurons can induce oxidative stress on the lipid content of the brain by increasing ROS and affecting antioxidant properties. QDs can break down and release metal ions, which can be carcinogenic and lead to inflammation and oxidative stress. QD-PEG-OMe, for example, can induce high expression of metallothionein (MT1) proteins and lysosomal dysfunction at a concentration as low as 0.1 μM and 1 μM can lead to cell death [147,155]. QDs accumulating in the spleen, thymus, and other immune organs can interact with immune cells, including macrophages, and potentially lead to cytotoxicity. CdSe/Zn QDs, at concentrations between 1.25 to 2.5 nm., have been shown to decrease cell proliferation and viability of macrophages, increase the production of ROS and apoptosis, and affect phagocytic activity and cytokine release. On the other hand, lymphocytes exposed to CdSe/Zn QDs exhibited enhanced cell viability and increased release of particular cytokines (IL-6 and TNF-α) [156,157]. QDs can also cause necrotic regions, increased lactate dehydrogenase (LDH) activity, and oxidative stress in the spleen and thymus [147].

QDs can reach the heart through the bloodstream and accumulate in valves, veins, and arterioles. In zebrafish studies (Fig. 9), increased concentrations of GQDs (up to 200 μg mL$^{-1}$) were observed to decrease heart rate without affecting blood flow, indicating potential bradycardia with higher QD concentrations [158]. Fig. 9 (Ref. [161]) shows the accumulation for QDs in different parts of the Zebrafish under brightfield and fluorescent microscopy. Fluorescence can be seen in different parts of the Zebrafish, including blood vessels, indicating the potential of QDs to accumulate in the heart and cause bradycardia. QDs can affect the reproductive system in both males and females. In male BALB/c mice, CdSe/ZnS QDs were found to impair spermatogenesis, reduce the number of spermatogonia, spermatids, and sperm, decrease overall testis weight, decrease testosterone levels, and increase luteinizing hormone (LH) levels. In female BALB/c mice, CdSe/ZnS QDs led to a downregulation of follicle-stimulating hormone (FSH) and LH and a decrease in the number of oocytes [159,160]. It is essential to consider these potential toxic effects and continue research to develop safer alternatives or mitigate the toxicity of QDs for biomedical applications.

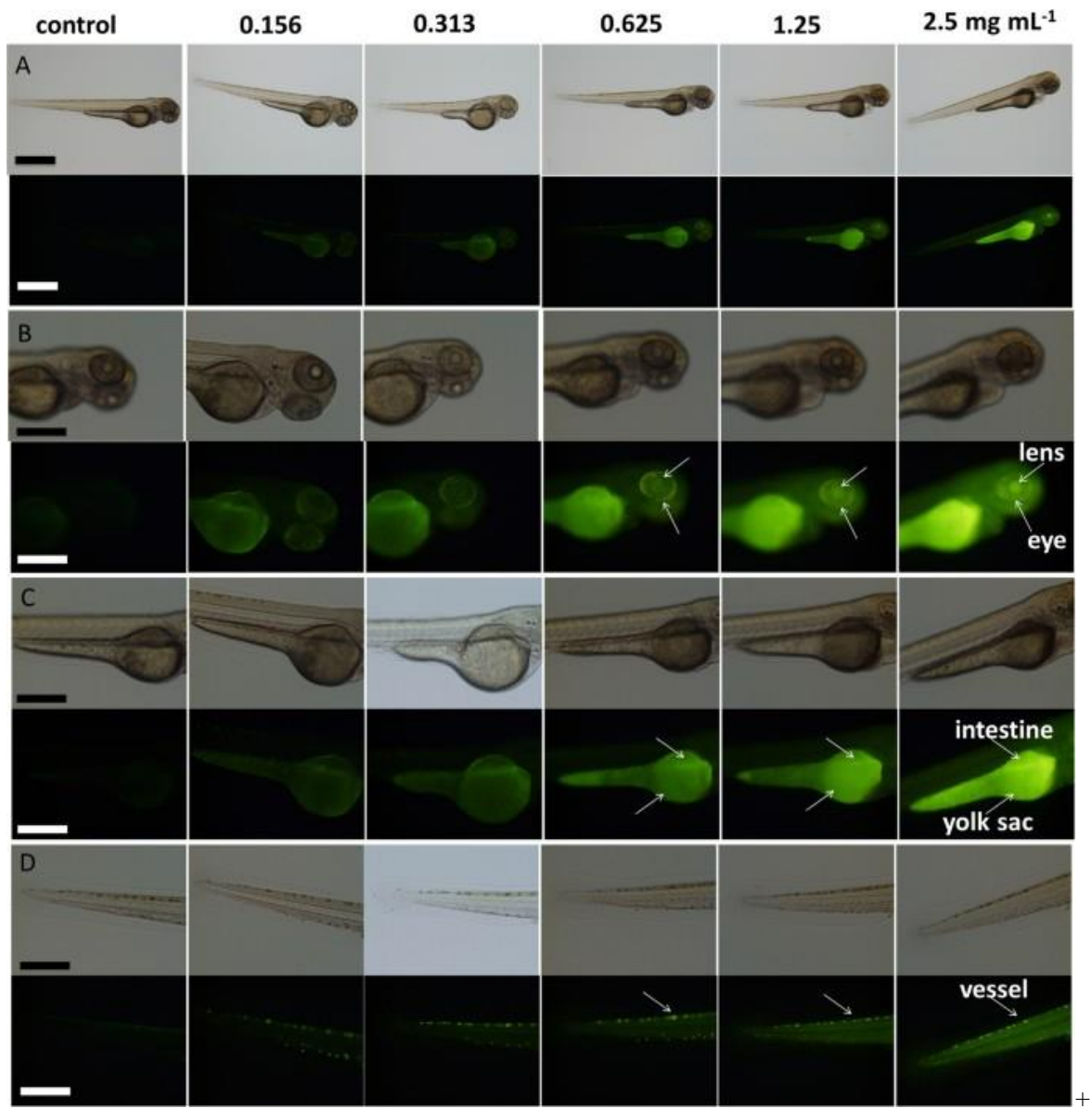

**Fig. 9. Accumulation of QDs in different organs of Zebrafish.** Brightfield (top) and Fluorescent (bottom) images of zebrafish at 84 hpf after soaking them in a C-QDs solution for 10 hours. (A) whole bodies, (B) head, (C) yolk sac, and (D) tail of zebrafish larvae. Reproduced with permission [161] Copyright 2015, Nature.

*Toxicity mitigation of QDs*

QDs have a wide range of applications in biomedicine. Therefore, enhancing their biocompatibility and reducing their toxicity to the host is crucial. Pdots are made up of organic compounds like PEG and PEI, which makes them comparatively less toxic to the host than other types of QDs. The weathering of QDs can result in toxicity; hence, a model has to be established to evaluate their safety. To address this, organic ligands such as humic acids and proteins reduce the bioavailability of metal ions (which are the primary cause of toxicity) [162]. Surface coatings or the addition of functional groups can make the QDs less toxic. In aquatic environments, extracellular polymeric substances derived from organisms like *Scenedesmus obliquus* form an "eco-corona" and can adhere to CdSe/ZnS QDs. This interaction modifies the physicochemical characteristics of the QDs, making them more biocompatible by reducing the effective surface area available for binding to host cells [163]. Another approach to reduce the toxicity of QDs, especially at high concentrations, is encapsulating GQDs within a PEG matrix. This encapsulation lowers the production of ROS and prevents cellular uptake [164,165]. NPs conjugated with titanium dioxide ($TiO_2$) have exhibited cytotoxic, carcinogenic, and increased ROS properties. N-GQDs can be adsorbed onto the surface of $TiO_2$ NPs, forming N-GQDs/$TiO_2$ nanocomposites to mitigate these issues [166]. The CdSe QDs release $Cd^{2+}$ ions, which lead to toxicity, but adding a ZnS shell reduces their cytotoxicity [167].

In addition to the aforementioned approaches, other strategies for enhancing the biocompatibility of QDs include using lower doses of QDs. A study conducted by Hoshino *et al*. [168] involving EL-4 cells incubated with CdSe/ZnS–SSA QDs (at the concentrations of 0.1, 0.2, and 0.4 mg/mL) revealed that as the concentration increased, the viability of the cells decreased, i.e., cytotoxicity increased. Similar studies were conducted by Tang *et al*. [169], wherein BALB/c mice were injected with concentrations 4.096, 5.12, 6.4, 8, to 10 nmol/kg of various QDs, and the mortality rate was assessed. At higher doses, the mortality rate of the mice was seen to increase [169]. That shows that higher doses of QDs can be toxic to cells. Modulating the size of QDs can also help in mitigating their toxicity. Lovrić *et al*. [170], conducted a study on PC12 and N9 cells treated with small green-emitting positively charged CdTe QDs and larger red-emitting CdTe QDs. The green-emitting QDs were seen to accumulate in the nucleus and organelles, whereas the red-emitting QDs were only found in the cytoplasm. That can explain how the size of QDs affects their accumulation in different organs, which can be customized according to the need [170]. Careful selection of materials can promote the biodegradability of QDs, making them more biocompatible. Furthermore, doping QDs reduces the likelihood of cytotoxicity and offers additional benefits such as avoiding self-quenching, increasing the band gap, and enabling fluorescent or chemo imaging [7].

## Challenges and Future Directions

QDs have gained significant attention in non-biological applications because of their small size and tunable electrical and optical properties. Their unique features make them a promising option in biomedicine compared to conventional organic dyes and fluorescent chemicals [171]. Their distinct optical properties make them an attractive alternative to traditional fluorophores. With advancements in synthesis techniques, it is now possible to tailor the size, surface functionality, bioconjugation, and targeting capabilities of biocompatible QDs, thus enabling the creation of a wide range of semiconductor nanocrystals [172]. However, several factors, including water solubility, stability, and toxicity, must be carefully considered when contemplating their utilization within biological environments [171].

### *Challenges in the Use of QDs in Biomedical Applications*

QDs in biological settings pose challenges due to their nano-colloidal behavior, which sets them apart from traditional organic fluorophores. That introduces safety concerns, including limited biological selectivity, complex surface chemistry, unregulated biodistribution to target tissues, limited water solubility, and the potential for long-term toxicity. Overcoming these challenges is crucial to direct QD biodistribution to specific organs, minimize long-term accumulation, and mitigate long-term toxicity. The fate of QDs *in vivo* depends on their chemistry and resulting physicochemical properties [171]. Despite significant progress and promising potential, QDs are currently limited to fundamental research because of long-term degradation, chronic toxicity, short-term stability, and biocompatibility issues. In addition to intrinsic physicochemical properties, cytotoxicity is influenced by environmental factors such as size, color, QD dosage, coating bioactivity, processing parameters, capping materials, and surface chemistry. The toxicity of QDs restricts their application in a limited range of *in-vivo* studies. Furthermore, the lack of homogeneity among designed QDs and their predominant use in bulk solutions hinders the development of reusable sensors and routine analytical applications. Prolonged exposure of CdSe QDs to UV light can lead to cell damage due to the production of $Cd^{2+}$ ions during photolysis. The fundamental aspects of QDs' physicochemical makeup and surface chemistry remain poorly understood in various contexts, raising questions about their potential application in point-of-care testing and requiring further investigation. Novel production techniques should be explored to minimize risks and ensure effective biological passivation of QDs. Enhancing stability in ambient conditions and investigating their shelf life are additional challenges that require attention [53].

### *Future Directions for QD Research in Biomedicine*

QDs have many clinical applications in neuroscience, including particle monitoring, cell labeling, drug administration, imaging, and photodynamic treatment. QDs hold promise for accurately detecting different types of cancer with minimal side effects, identifying molecular causes of diseases, and discovering potential therapeutic targets. They can be used for intracellular and extracellular research, and the development of novel biochemical testing approaches. Ongoing research aims to enhance QD efficiency and fluorescence by investigating more luminous hydrophilic QDs. Scientific modifications have successfully reduced the biotoxicity and bioaccumulation of QDs. However, balancing reducing hazardous components and maintaining desired optical properties remains challenging. Excellent biocompatibility ensures the widespread use of QDs in botanical applications, such as biology and biomedicine. QDs and other innovative nanomaterials serve as valuable research tools for studying plants, and understanding the ecological toxicity of nanomaterials is essential [173,174]. In biomedicine, QDs find applications in cellular probes, cell tracking, intracellular drug delivery, radio-opacity, paramagnetic properties,

MRI contrast agents, bio-imaging, tissue staining, gene delivery, cellular imaging, and cellular motility assays. QDs GQDs have emerged as a promising biomedical tool due to their large surface area and functional groups, which enable easy conjugation with various drugs and ligands for targeted delivery [175]. Zero-dimensional black phosphorous QDs (BPQD) has also gained attention in biological imaging, drug administration, and photothermal treatment [176]. These advancements and potential future applications highlight the immense possibilities QDs offer in various biomedicine fields. Thorough research and exploration of QDs can lead to remarkable breakthroughs and advancements in the future.

## Conclusion

QDs possess unique properties that make them promising tools for biomedical applications, particularly in bioimaging, drug delivery, sensing and detection, photodynamic therapy, and tissue engineering. Their strong fluorescence, photostability, and ability to bind to biomolecules make them excellent for detecting analytes and assessing drug efficacy. In tissue engineering, QDs serve as bone and cartilage regeneration scaffolds, enhancing cell attachment, proliferation, and differentiation. However, the toxicity and biocompatibility of QDs must be carefully evaluated before their clinical use. QDs also show great potential in multimodal imaging techniques, nanocarrier-mediated drug delivery, and early detection of cancer cells, among other applications. Surface modifications, encapsulation techniques, and biodegradable materials can be further explored to reduce QD toxicity and enhance biocompatibility. Larger QDs can be more toxic due to difficulties in their excretion, potentially causing damage to organs such as the liver, lungs, kidneys, brain, and immune system. Further studies are needed to understand QD toxicity and its implications thoroughly. Ongoing research and development efforts aim to unlock the full potential of QDs in various biomedical applications. Despite the challenges posed by their toxicity, QDs offer significant promise in biomedicine. With continued advancements and careful exploration of their properties, QDs have the potential to revolutionize biomedical applications and contribute to advancements in healthcare.

## Author Contributions
PB: Conceptualization, Writing— Review and Editing, Supervision, Funding acquisition. AS, AP, and BS: Writing— Original draft preparation, visualization. ADBM: Writing— Review and editing, Funding acquisition.

## Ethics Approval and Consent to Participate
Not applicable.

## Acknowledgment
Not applicable.

## Funding
This work was supported by Dayananda Sagar University Seed Grant (Grant No. DSU/REG/2022-23/740).

## Conflict of Interest
The authors declare that they have no known competing financial interests or personal relationships that could have appeared to influence the work reported in this paper.

## References
[1] Matea CT, Mocan T, Tabaran F, Pop T, Mosteanu O, Puia C, *et al*. Quantum dots in imaging, drug delivery and sensor applications. International Journal of Nanomedicine. 2017; 12: 5421–5431.
[2] Badıllı U, Mollarasouli F, Bakirhan NK, Ozkan Y, Ozkan SA. Role of quantum dots in pharmaceutical and biomedical analysis, and its application in drug delivery. TrAC Trends in Analytical Chemistry. 2020; 131: 116013.
[3] Martynenko IV, Litvin AP, Purcell-Milton F, Baranov AV, Fedorov AV, Gun'ko YK. Application of semiconductor quantum dots in bioimaging and biosensing. Journal of Materials Chemistry. B. 2017; 5: 6701–6727.
[4] Pleskova S, Mikheeva E, Gornostaeva E. Using of Quantum Dots in Biology and Medicine. Advances in Experimental Medicine and Biology. 2018; 1048: 323–334.
[5] Smyder JA, Krauss TD. Coming attractions for semiconductor quantum dots. Materials Today. 2011; 14: 382–387
[6] Gidwani B, Sahu V, Shukla SS, Pandey R, Joshi V, Jain VK, *et al*. Quantum dots: Prospectives, toxicity, advances and applications. Journal of Drug Delivery Science and Technology. 2021; 61: 102308.
[7] Wu P, Yan XP. Doped quantum dots for chemo/biosensing and bioimaging. Chemical Society Reviews. 2013; 42: 5489–5521.
[8] Wang L, Xu D, Gao J, Chen X, Duo Y, Zhang H. Semiconducting quantum dots: Modification and applications in biomedical science. Science China Materials. 2020; 63: 1631–1650.
[9] Wu X, Tian F, Zhao JX, Wu M. Evaluating pharmacokinetics and toxicity of luminescent quantum dots. Expert Opinion on Drug Metabolism & Toxicology. 2013; 9: 1265–1277.

[10] Kargozar S, Hoseini SJ, Milan PB, Hooshmand S, Kim HW, Mozafari M. Quantum Dots: A Review from Concept to Clinic. Biotechnology Journal. 2020; 15: e2000117.

[11] Hezinger AFE, Tessmar J, Göpferich A. Polymer coating of quantum dots--a powerful tool toward diagnostics and sensorics. European Journal of Pharmaceutics and Biopharmaceutics: Official Journal of Arbeitsgemeinschaft Fur Pharmazeutische Verfahrenstechnik E.V. 2008; 68: 138–152.

[12] David A, Burkard G, Kormányos A. Effective theory of monolayer TMDC double quantum dots. 2D Mater. 2018; 5: 035031.

[13] Sharbirin AS, Akhtar S, Kim J. Light-emitting MXene quantum dots. Opto-Electronic Advances. 2021; 4: 200077–200077.

[14] Wang HC, Bao Z, Tsai HY, Tang AC, Liu RS. Perovskite Quantum Dots and Their Application in Light-Emitting Diodes. Small (Weinheim an Der Bergstrasse, Germany). 2018; 14: 10.1002/smll.201702433.

[15] Maxwell T, Nogueira Campos MG, Smith S, Doomra M, Thwin Z, Santra S. Quantum Dots Nanoparticles for Biomedical Applications (pp 243–65). Elsevier: Location. 2020

[16] Yu P, Wang ZM. Quantum Dot Optoelectronic Devices. Springer International Publishing: Cham. 2020.

[17] Chang C-Y, Venkatesan S, Herman A, Wang C-L, Teng H, Lee Y-L. Carbon quantum dots with high quantum yield prepared by heterogeneous nucleation processes. Journal of Alloys and Compounds. 2023 938 168654.

[18] Fitzpatrick JAJ, Andreko SK, Ernst LA, Waggoner AS, Ballou B, Bruchez MP. Long-term persistence and spectral blue shifting of quantum dots *in vivo*. Nano Letters. 2009; 9: 2736–2741.

[19] Jha S, Mathur P, Ramteke S, Jain NK. Pharmaceutical potential of quantum dots. Artificial Cells, Nanomedicine, and Biotechnology. 2018; 46: 57–65.

[20] Hu J, Zhang CY. Simple and accurate quantification of quantum yield at the single-molecule/particle level. Analytical Chemistry. 2013; 85: 2000–2004.

[21] Yang W, Yang H, Ding W, Zhang B, Zhang L, Wang L, *et al*. High quantum yield ZnO quantum dots synthesizing via an ultrasonication microreactor method. Ultrasonics Sonochemistry. 2016; 33: 106–117.

[22] Tan X, Li Y, Li X, Zhou S, Fan L, Yang S. Electrochemical synthesis of small-sized red fluorescent graphene quantum dots as a bioimaging platform. Chemical Communications (Cambridge, England). 2015; 51: 2544–2546.

[23] Lee KH, Lee JH, Kang HD, Park B, Kwon Y, Ko H, *et al*. Over 40 cd/A efficient green quantum dot electroluminescent device comprising uniquely large-sized quantum dots. ACS Nano. 2014; 8: 4893–4901.

[24] Cai W, Hsu AR, Li ZB, Chen X. Are quantum dots ready for *in vivo* imaging in human subjects? Nanoscale Research Letters. 2007; 2: 265–281.

[25] Iannazzo D, Ziccarelli I, Pistone A. Graphene quantum dots: multifunctional nanoplatforms for anticancer therapy. Journal of Materials Chemistry B. 2017; 5: 6471–6489.

[26] Hasanirokh K, Asgari A, Mohammadi S. Fabrication of a light-emitting device based on the CdS/ZnS spherical quantum dots. Journal of the European Optical Society-Rapid Publications. 2021; 17: 26.

[27] Jiang Y, Cho S-Y, Shim M. Light-emitting diodes of colloidal quantum dots and nanorod heterostructures for future emissive displays. Journal of Materials Chemistry C. 2018; 6: 2618–2634.

[28] Kim T, Kang H, Jeong S, Kang DJ, Lee C, Lee CH, *et al*. Au@polymer core-shell nanoparticles for simultaneously enhancing efficiency and ambient stability of organic optoelectronic devices. ACS Applied Materials & Interfaces. 2014; 6: 16956–16965.

[29] Brkić S. Applicability of Quantum Dots in Biomedical Science Ionizing Radiation Effects and Applications. In Djezzar B (ed.) InTech: Rijeka. 2018.

[30] Rizvi SB, Ghaderi S, Keshtgar M, Seifalian AM. Semiconductor quantum dots as fluorescent probes for *in vitro* and *in vivo* bio-molecular and cellular imaging. Nano Reviews. 2010; 1: 10.3402/nano.v1i0.5161.

[31] Jenjob R, Phakkeeree T, Crespy D. Core-shell particles for drug-delivery, bioimaging, sensing, and tissue engineering. Biomaterials Science. 2020; 8: 2756–2770.

[32] Goswami PN, Mandal D, Rath AK. The role of surface ligands in determining the electronic properties of quantum dot solids and their impact on photovoltaic figure of merits. Nanoscale. 2018; 10: 1072–1080.

[33] Paez-Ornelas JI, Ponce-Pérez R, Fernández-Escamilla HN, Hoat DM, Murillo-Bracamontes EA, Moreno-Armenta MG, *et al*. The effect of shape and size in the stability of triangular Janus MoSSe quantum dots. Scientific Reports. 2021; 11: 21061.

[34] Sahu J, Prusty D, Mansingh S, Parida K. A review on alloyed quantum dots and their applications as photocatalysts. International Journal of Hydrogen Energy. 2023; 48: 29097–29118.

[35] Hu R, Yong KT, Roy I, Ding H, Law WC, Cai H, *et al*. Functionalized near-infrared quantum dots for in vivo tumor vasculature imaging. Nanotechnology. 2010; 21: 145105.

[36] Misra R, Das M, Biswas P, Nanda A. EGFR targeted Mn-doped ZnO fluorescent nanocrystals for cancer theranostic application. Materials Today Communications. 2021; 26: 102170.

[37] Khoshnood A, Farhadian N, Abnous K, Matin M M, Ziaee N, Yaghoobi E. N doped-carbon quantum dots with ultra-high quantum yield photoluminescent property conjugated with folic acid for targeted drug delivery and bioimaging applications. Journal of Photochemistry and Photobiology A: Chemistry. 2023; 444: 114972.

[38] Shamsipur M, Reza Rajabi H, Khani O. Pure and $Fe^{3+}$-doped ZnS quantum dots as novel and efficient nanophotocatalysts: Synthesis, characterization and use for decolorization of Victoria blue R. Materials Science in Semiconductor Processing. 2013; 16: 1154–1161.

[39] Li Y, Ding Y, Zhang Y, Qian Y. Photophysical properties of ZnS quantum dots Journal of Physics and Chemistry of Solids. 1999; 60: 13–15.

[40] Sharma K, Raizada P, Hasija V, Singh P, Bajpai A, Nguyen V-H, *et al*. ZnS-based quantum dots as photocatalysts for water purification. Journal of Water Process Engineering. 2021; 43: 102217.

[41] Jana P, Dev A. Carbon quantum dots: A promising nanocarrier for bioimaging and drug delivery in cancer. Materials Today Communications. 2022; 32: 104068.

[42] Hasan AMM, Hasan Md A, Reza A, Islam Md M, Susan Md ABH. Carbon dots as nano-modules for energy conversion and storage. Materials Today Communications. 2021; 29: 102732.

[43] Bacon M, Bradley SJ, Nann T. Graphene Quantum Dots. Particle & Particle Systems Characterization. 2014; 31: 415–428.

[44] Kundu S, Pillai VK. Synthesis and characterization of graphene quantum dots. Physical Sciences Reviews. 2019; 5.

[45] Yang Y, Chen S, Liu L, Li S, Zeng Q, Zhao X, *et al*. Increasing Cancer Therapy Efficiency through Targeting and Localized Light Activation. ACS Applied Materials & Interfaces. 2017; 9: 23400–23408.

[46] Chen F, Gao W, Qiu X, Zhang H, Liu L, Liao P, et al. Graphene quantum dots in biomedical applications: Recent advances and future challenges Frontiers in Laboratory Medicine. 2017; 1: 192–199.
[47] Ding K, Jing L, Liu C, Hou Y, Gao M. Magnetically engineered Cd-free quantum dots as dual-modality probes for fluorescence/magnetic resonance imaging of tumors. Biomaterials. 2014; 35: 1608–1617.
[48] Jing L, Ding K, Kershaw SV, Kempson IM, Rogach AL, Gao M. Magnetically engineered semiconductor quantum dots as multimodal imaging probes. Advanced Materials (Deerfield Beach, Fla.). 2014; 26: 6367–6386.
[49] Shi W, Han Q, Wu J, Ji C, Zhou Y, Li S, et al. Synthesis Mechanisms, Structural Models, and Photothermal Therapy Applications of Top-Down Carbon Dots from Carbon Powder, Graphite, Graphene, and Carbon Nanotubes. International Journal of Molecular Sciences. 2022; 23: 1456.
[50] Choi Y, Choi Y, Kwon OH, Kim BS. Carbon Dots: Bottom-Up Syntheses, Properties, and Light-Harvesting Applications. Chemistry, an Asian Journal. 2018; 13: 586–598.
[51] Joglekar PV, Mandalkar DJ, Nikam MA, Pande NS, Dubal A. Review article on Quantum Dots: Synthesis, Properties and Application. International Journal of Research in Advent Technology. 2019; 7: 510–515.
[52] Valizadeh A, Mikaeili H, Samiei M, Farkhani SM, Zarghami N, Kouhi M, et al. Quantum dots: synthesis, bioapplications, and toxicity. Nanoscale Research Letters. 2012; 7: 480.
[53] Singh S, Dhawan A, Karhana S, Bhat M, Dinda AK. Quantum Dots: An Emerging Tool for Point-of-Care Testing. Micromachines. 2020; 11: 1058.
[54] Pillar-Little TJ, Wanninayake N, Nease L, Heidary DK, Glazer EC, Kim DY. Superior photodynamic effect of carbon quantum dots through both type I and type II pathways: Detailed comparison study of top-down-synthesized and bottom-up-synthesized carbon quantum dots. Carbon. 2018; 140: 616–623.
[55] Hosseini MS, Kamali M. Synthesis and characterization of aspartic acid-capped CdS/ZnS quantum dots in reverse micelles and its application to Hg(II) determination. Journal of Luminescence. 2015; 167: 51–58.
[56] Rebohle L, Wutzler R, Prucnal S, Hübner R, Georgiev YM, Erbe A, et al. Local Formation of InAs Nanocrystals in Si by Masked Ion Implantation and Flash Lamp Annealing. Physica Status Solidi C. 2017; 14:
[57] Chee SW, Kammler M, Graham J, Gignac L, Reuter MC, Hull R, et al. Directed Self-Assembly of Ge Quantum Dots Using Focused $Si^{2+}$ Ion Beam Patterning. Scientific Reports. 2018; 8: 9361.
[58] Sun L, Wang W, Yang H. Recent Progress in Synaptic Devices Based on 2D Materials. Advanced Intelligent Systems. 2020; 2.
[59] Harmon KJ, Delegan N, Highland MJ, He H, Zapol P, Heremans FJ, et al. Designing silicon carbide heterostructures for quantum information science: challenges and opportunities. Materials for Quantum Technology. 2022; 2: 023001.
[60] Perez-Roldan M, Colpo P, Gilliland D, Ceccone G, Rossi F. Surface characterisation of PEO-like microstructures by means of ToF-SIMS, XPS and SPR. Surface and Interface Analysis. 2013; 45: 240–243.
[61] Grella L, Carroll A, Murray K, McCord MA, Tong WM, Brodie AD, et al. Digital pattern generator: an electron-optical MEMS for massively parallel reflective electron beam lithography. Journal of Micro/Nanolithography, MEMS, and MOEMS. 2013; 12: 031107.
[62] Heuberger A. X-ray lithography. Microelectronic Engineering. 1986; 5: 3–38.
[63] Sharma E, Rathi R, Misharwal J, Sinhmar B, Kumari S, Dalal J, et al. Evolution in Lithography Techniques: Microlithography to Nanolithography. Nanomaterials (Basel, Switzerland). 2022; 12: 2754.
[64] Bera D, Qian L, Tseng T-K, Holloway PH. Quantum Dots and Their Multimodal Applications: A Review. Materials. 2010; 3: 2260–2345.
[65] Ling D, Hackett MJ, Hyeon T. Surface ligands in synthesis, modification, assembly and biomedical applications of nanoparticles Nano Today. 2014; 9: 457–477.
[66] Fernández-Argüelles MT, Jin WJ, Costa-Fernández JM, Pereiro R, Sanz-Medel A. Surface-modified CdSe quantum dots for the sensitive and selective determination of Cu(II) in aqueous solutions by luminescent measurements. Analytica Chimica Acta. 2005; 549: 20–25.
[67] Zhang B, Hu R, Wang Y, Yang C, Liu X, Yong K-T. Revisiting the principles of preparing aqueous quantum dots for biological applications: the effects of surface ligands on the physicochemical properties of quantum dots. RSC Advances. 2014; 4: 13805–13816.
[68] Silva FO, Carvalho MS, Mendonça R, Macedo WA, Balzuweit K, Reiss P, et al. Effect of surface ligands on the optical properties of aqueous soluble CdTe quantum dots. Nanoscale Research Letters. 2012; 7: 536.
[69] Drozd D, Zhang H, Goryacheva I, De Saeger S, Beloglazova NV. Silanization of quantum dots: Challenges and perspectives. Talanta. 2019; 205: 120164.
[70] Ma N, Marshall AF, Gambhir SS, Rao J. Facile synthesis, silanization, and biodistribution of biocompatible quantum dots. Small (Weinheim an Der Bergstrasse, Germany). 2010; 6: 1520–1528.
[71] Liu X, Luo Y. Surface Modifications Technology of Quantum Dots Based Biosensors and Their Medical Applications Chinese Journal of Analytical Chemistry. 2014; 42: 1061–1069.
[72] Alejo T, Merchán MD, Velázquez MM. Adsorption of quantum dots onto polymer and Gemini surfactant films: a quartz crystal microbalance study. Langmuir: the ACS Journal of Surfaces and Colloids. 2014; 30: 9977–9984.
[73] Park J, Lee J, Kwag J, Baek Y, Kim B, Yoon CJ, et al. Quantum Dots in an Amphiphilic Polyethyleneimine Derivative Platform for Cellular Labeling, Targeting, Gene Delivery, and Ratiometric Oxygen Sensing. ACS Nano. 2015; 9: 6511–6521.
[74] Nie X, Zhang Y, Wang X, Ren C, Gao S-Q, Lin Y-W. Direct Visualization of Ligands Exchange on the Surfaces of Quantum Dots by a Two-Phase Approach. ChemistrySelect. 2018; 3: 2267–2271.
[75] Ma Y, Li Y, Ma S, Zhong X. Highly bright water-soluble silica coated quantum dots with excellent stability. Journal of Materials Chemistry. B. 2014; 2: 5043–5051.
[76] Lee J, Im JH, Huh KM, Lee YK, Shin H. Preparation and characterization of CdSe/ZnS quantum dots encapsulated in poly(ethylene glycol)-b-poly(D,L-lactide) micelle nanoparticles. Journal of Nanoscience and Nanotechnology. 2010; 10: 487–496.
[77] John BK, Abraham T, Mathew B. A Review on Characterization Techniques for Carbon Quantum Dots and Their Applications in Agrochemical Residue Detection. Journal of Fluorescence. 2022; 32: 449–471.
[78] Hessman D, Castrillo P, Pistol M-E, Pryor C, Samuelson L. Excited states of individual quantum dots studied by photoluminescence spectroscopy. Applied Physics Letters. 1996; 69: 749–751.

[79] Tabish TA, Lin L, Ali M, Jabeen F, Ali M, Iqbal R, *et al*. Investigating the bioavailability of graphene quantum dots in lung tissues via Fourier transform infrared spectroscopy. Interface Focus. 2018; 8: 20170054.
[80] Hazdra P, Voves J, Oswald J, Kuldová K, Hospodková A, Hulicius E, *et al*. Optical characterisation of MOVPE grown vertically correlated InAs/GaAs quantum dots. Microelectronics Journal. 2008; 39: 1070–1074.
[81] Gu Y, Kuskovsky IL, Fung J, Robinson R, Herman IP, Neumark GF, *et al.* Determination of size and composition of optically active CdZnSe/ZnBeSe quantum dots. Applied Physics Letters. 2003; 83; 3779–3781.
[82] Doty MF, Climente JI, Korkusinski M, Scheibner M, Bracker AS, Hawrylak P, *et al*. Antibonding ground states in InAs quantum-dot molecules. Physical Review Letters. 2009; 102: 047401.
[83] Lees EE, Gunzburg MJ, Nguyen TL, Howlett GJ, Rothacker J, Nice EC, *et al*. Experimental determination of quantum dot size distributions, ligand packing densities, and bioconjugation using analytical ultracentrifugation. Nano Letters. 2008; 8: 2883–2890.
[84] Drbohlavova J, Adam V, Kizek R, Hubalek J. Quantum dots - characterization, preparation and usage in biological systems. International Journal of Molecular Sciences. 2009; 10: 656–673.
[85] Gao X, Yang L, Petros JA, Marshall FF, Simons JW, Nie S. In vivo molecular and cellular imaging with quantum dots. Current Opinion in Biotechnology. 2005; 16: 63–72.
[86] Li J, Zhu JJ. Quantum dots for fluorescent biosensing and bio-imaging applications. The Analyst. 2013; 138: 2506–2515.
[87] Ji X, Peng F, Zhong Y, Su Y, He Y. Fluorescent quantum dots: synthesis, biomedical optical imaging, and biosafety assessment. Colloids and Surfaces. B, Biointerfaces. 2014; 124: 132–139.
[88] Shi X, Meng X, Sun L, Liu J, Zheng J, Gai H, *et al*. Observing photophysical properties of quantum dots in air at the single molecule level: advantages in microarray applications. Lab on a Chip. 2010; 10: 2844–2847.
[89] Li H, Ye S, Guo J, Wang H, Yan W, Song J, *et al*. Biocompatible carbon dots with low-saturation-intensity and high-photobleaching-resistance for STED nanoscopy imaging of the nucleolus and tunneling nanotubes in living cells. Nano Research. 2019; 12: 3075–3084.
[90] Liang Z, Khawar MB, Liang J, Sun H. Bio-Conjugated Quantum Dots for Cancer Research: Detection and Imaging. Frontiers in Oncology. 2021; 11: 749970.
[91] Wu SM, Zhao X, Zhang ZL, Xie HY, Tian ZQ, Peng J, *et al*. Quantum-dot-labeled DNA probes for fluorescence *in situ* hybridization (FISH) in the microorganism Escherichia coli. Chemphyschem: a European Journal of Chemical Physics and Physical Chemistry. 2006; 7: 1062–1067.
[92] Walling MA, Novak JA, Shepard JRE. Quantum dots for live cell and in vivo imaging. International Journal of Molecular Sciences. 2009; 10: 441–491.
[93] Gao X, Chung LWK, Nie S. Quantum Dots for In Vivo Molecular and Cellular Imaging. In Bruchez MP, Hotz CZ (eds.) Quantum Dots (pp 135–146). Humana Press: New Jersey. 2007.
[94] Tao H, Yang K, Ma Z, Wan J, Zhang Y, Kang Z, *et al*. In vivo NIR fluorescence imaging, biodistribution, and toxicology of photoluminescent carbon dots produced from carbon nanotubes and graphite. Small (Weinheim an Der Bergstrasse, Germany). 2012; 8: 281–290.
[95] Yang W, Guo W, Gong X, Zhang B, Wang S, Chen N, *et al*. Facile Synthesis of Gd-Cu-In-S/ZnS Bimodal Quantum Dots with Optimized Properties for Tumor Targeted Fluorescence/MR In Vivo Imaging. ACS Applied Materials & Interfaces. 2015; 7: 18759–18768.
[96] Koole R, Mulder WJM, van Schooneveld MM, Strijkers GJ, Meijerink A, Nicolay K. Magnetic quantum dots for multimodal imaging. Wiley Interdisciplinary Reviews. Nanomedicine and Nanobiotechnology. 2009; 1: 475–491.
[97] Mulder WJM, Strijkers GJ, Nicolay K, Griffioen AW. Quantum dots for multimodal molecular imaging of angiogenesis. Angiogenesis. 2010; 13: 131–134.
[98] Bouzas-Ramos D, Cigales Canga J, Mayo JC, Sainz RM, Ruiz Encinar J, Costa-Fernandez JM. Carbon Quantum Dots Codoped with Nitrogen and Lanthanides for Multimodal Imaging. Advanced Functional Materials. 2019; 29: 1903884.
[99] Zhao Y, Hao X, Lu W, Wang R, Shan X, Chen Q, *et al*. Facile Preparation of Double Rare Earth-Doped Carbon Dots for MRI/CT/FI Multimodal Imaging. ACS Applied Nano Materials - ACS Publications. 2018; 1; 2544–2551.
[100] Wang S, Jarrett BR, Kauzlarich SM, Louie AY. Core/shell quantum dots with high relaxivity and photoluminescence for multimodality imaging. Journal of the American Chemical Society. 2007; 129: 3848–3856.
[101] Zhao JY, Chen G, Gu YP, Cui R, Zhang ZL, Yu ZL, *et al*. Ultrasmall Magnetically Engineered Ag2Se Quantum Dots for Instant Efficient Labeling and Whole-Body High-Resolution Multimodal Real-Time Tracking of Cell-Derived Microvesicles. Journal of the American Chemical Society. 2016; 138: 1893–1903.
[102] Wang Y, Wu B, Yang C, Liu M, Sum TC, Yong KT. Synthesis and Characterization of Mn:ZnSe/ZnS/ZnMnS Sandwiched QDs for Multimodal Imaging and Theranostic Applications. Small (Weinheim an Der Bergstrasse, Germany). 2016; 12: 534–546.
[103] Probst CE, Zrazhevskiy P, Bagalkot V, Gao X. Quantum dots as a platform for nanoparticle drug delivery vehicle design. Advanced Drug Delivery Reviews. 2013; 65: 703–718.
[104] Qi L, Gao X. Emerging application of quantum dots for drug delivery and therapy. Expert Opinion on Drug Delivery. 2008; 5: 263–267.
[105] Michalet X, Pinaud FF, Bentolila LA, Tsay JM, Doose S, Li JJ, *et al*. Quantum dots for live cells, in vivo imaging, and diagnostics. Science (New York, N.Y.). 2005; 307: 538–544.
[106] Shahbazi R, Ozpolat B, Ulubayram K. Oligonucleotide-based theranostic nanoparticles in cancer therapy. Nanomedicine (London, England). 2016; 11: 1287–1308.
[107] Ho YP, Leong KW. Quantum dot-based theranostics. Nanoscale. 2010; 2: 60–68.
[108] Singh SP. Multifunctional magnetic quantum dots for cancer theranostics. Journal of Biomedical Nanotechnology. 2011; 7: 95–97.
[109] AbdElhamid AS, Helmy MW, Ebrahim SM, Bahey-El-Din M, Zayed DG, Zein El Dein EA, *et al*. Layer-by-layer gelatin/chondroitin quantum dots-based nanotheranostics: combined rapamycin/celecoxib delivery and cancer imaging. Nanomedicine (London, England). 2018; 13: 1707–1730.
[110] AbdElhamid AS, Zayed DG, Helmy MW, Ebrahim SM, Bahey-El-Din M, Zein-El-Dein EA, *et al*. Lactoferrin-tagged quantum dots-based theranostic nanocapsules for combined COX-2 inhibitor/herbal therapy of breast cancer. Nanomedicine (London, England). 2018; 13: 2637–2656.

[111] Wu C, Wang C, Han T, Zhou X, Guo S, Zhang J. Insight into the cellular internalization and cytotoxicity of graphene quantum dots. Advanced Healthcare Materials. 2013; 2: 1613–1619.
[112] Zhao C, Song X, Liu Y, Fu Y, Ye L, Wang N, et al. Synthesis of graphene quantum dots and their applications in drug delivery. Journal of Nanobiotechnology. 2020; 18: 142.
[113] Iannazzo D, Pistone A, Salamò M, Galvagno S, Romeo R, Giofré SV, et al. Graphene quantum dots for cancer targeted drug delivery. International Journal of Pharmaceutics. 2017; 518: 185–192.
[114] Nair A, Haponiuk JT, Thomas S, Gopi S. Natural carbon-based quantum dots and their applications in drug delivery: A review. Biomedicine & Pharmacotherapy = Biomedecine & Pharmacotherapie. 2020; 132: 110834.
[115] Ge J, Jia Q, Liu W, Lan M, Zhou B, Guo L, et al. Carbon Dots with Intrinsic Theranostic Properties for Bioimaging, Red-Light-Triggered Photodynamic/Photothermal Simultaneous Therapy In Vitro and In Vivo. Advanced Healthcare Materials. 2016; 5: 665–675.
[116] Yan F, Li L, Deng Z, Jin Q, Chen J, Yang W, et al. Paclitaxel-liposome-microbubble complexes as ultrasound-triggered therapeutic drug delivery carriers. Journal of Controlled Release: Official Journal of the Controlled Release Society. 2013; 166: 246–255.
[117] Liu Y, Zhang B, Yan B. Enabling anticancer therapeutics by nanoparticle carriers: the delivery of Paclitaxel. International Journal of Molecular Sciences. 2011; 12: 4395–4413.
[118] Zavvar T, Babaei M, Abnous K, Taghdisi SM, Nekooei S, Ramezani M, et al. Synthesis of multimodal polymersomes for targeted drug delivery and MR/fluorescence imaging in metastatic breast cancer model. International Journal of Pharmaceutics. 2020; 578: 119091.
[119] Zayed DG, AbdElhamid AS, Freag MS, Elzoghby AO. Hybrid quantum dot-based theranostic nanomedicines for tumor-targeted drug delivery and cancer imaging. Nanomedicine (London, England). 2019; 14: 225–228.
[120] Cai X, Luo Y, Zhang W, Du D, Lin Y. pH-Sensitive ZnO Quantum Dots-Doxorubicin Nanoparticles for Lung Cancer Targeted Drug Delivery. ACS Applied Materials & Interfaces. 2016; 8: 22442–22450.
[121] Gu YJ, Cheng J, Man CWY, Wong WT, Cheng SH. Gold-doxorubicin nanoconjugates for overcoming multidrug resistance. Nanomedicine: Nanotechnology, Biology, and Medicine. 2012; 8: 204–211.
[122] Ruzycka-Ayoush M, Kowalik P, Kowalczyk A, Bujak P, Nowicka AM, Wojewodzka M, et al. Quantum dots as targeted doxorubicin drug delivery nanosystems in human lung cancer cells. Cancer Nanotechnology. 2021; 12: 8.
[123] Shi J, Liu Y, Wang L, Gao J, Zhang J, Yu X, et al. A tumoral acidic pH-responsive drug delivery system based on a novel photosensitizer (fullerene) for in vitro and in vivo chemo-photodynamic therapy. Acta Biomaterialia. 2014; 10: 1280–1291.
[124] Liu L, Jiang H, Dong J, Zhang W, Dang G, Yang M, et al. PEGylated MoS$_2$ quantum dots for traceable and pH-responsive chemotherapeutic drug delivery. Colloids and Surfaces. B, Biointerfaces. 2020; 185: 110590.
[125] Choi HS, Liu W, Misra P, Tanaka E, Zimmer JP, Itty Ipe B, et al. Renal clearance of quantum dots. Nature Biotechnology. 2007; 25: 1165–1170.
[126] Hagens WI, Oomen AG, de Jong WH, Cassee FR, Sips AJAM. What do we (need to) know about the kinetic properties of nanoparticles in the body? Regulatory Toxicology and Pharmacology: RTP. 2007; 49: 217–229.
[127] Oberdörster G, Sharp Z, Atudorei V, Elder A, Gelein R, Lunts A, et al. Extrapulmonary translocation of ultrafine carbon particles following whole-body inhalation exposure of rats. Journal of Toxicology and Environmental Health. Part a. 2002; 65: 1531–1543.
[128] Hardman R. A toxicologic review of quantum dots: toxicity depends on physicochemical and environmental factors. Environmental Health Perspectives. 2006; 114: 165–172.
[129] Ballou B, Lagerholm BC, Ernst LA, Bruchez MP, Waggoner AS. Noninvasive imaging of quantum dots in mice. Bioconjugate Chemistry. 2004; 15: 79–86.
[130] Wang Y, Hu R, Lin G, Roy I, Yong KT. Functionalized quantum dots for biosensing and bioimaging and concerns on toxicity. ACS Applied Materials & Interfaces. 2013; 5: 2786–2799.
[131] Kansara V, Shukla R, Flora SJS, Bahadur P, Tiwari S. Graphene quantum dots: Synthesis, optical properties and navigational applications against cancer. Materials Today Communications. 2022; 31; 103359.
[132] Hai X, Feng J, Chen X, Wang J. Tuning the optical properties of graphene quantum dots for biosensing and bioimaging. Journal of Materials Chemistry. B. 2018; 6: 3219–3234.
[133] Pourmadadi M, Rahmani E, Rajabzadeh-Khosroshahi M, Samadi A, Behzadmehr R, Rahdar A, et al. Properties and application of carbon quantum dots (CQDs) in biosensors for disease detection: A comprehensive review. Journal of Drug Delivery Science and Technology. 2023; 80: 104156.
[134] Shen J, Zhou Y, Fu F, Xu H, Lv J, Xiong Y, et al. Immunochromatographic assay for quantitative and sensitive detection of hepatitis B virus surface antigen using highly luminescent quantum dot-beads. Talanta. 2015; 142: 145–149.
[135] Cui F, Ye Y, Ping J, Sun X. Carbon dots: Current advances in pathogenic bacteria monitoring and prospect applications. Biosensors & Bioelectronics. 2020; 156: 112085.
[136] Han G, Zhao J, Zhang R, Tian X, Liu Z, Wang A, et al. Membrane-Penetrating Carbon Quantum Dots for Imaging Nucleic Acid Structures in Live Organisms. Angewandte Chemie (International Ed. in English). 2019; 58: 7087–7091.
[137] Zhou JW, Zou XM, Song SH, Chen GH. Quantum Dots Applied to Methodology on Detection of Pesticide and Veterinary Drug Residues. Journal of Agricultural and Food Chemistry. 2018; 66: 1307–1319.
[138] Cardoso Dos Santos M, Algar WR, Medintz IL, Hildebrandt N. Quantum dots for Förster Resonance Energy Transfer (FRET) TrAC. Trends in Analytical Chemistry. 2020; 125: 115819.
[139] León-Simón G, Lazcano Z, Pérez E, Meza O. Förster resonance energy transfer in finite length systems and porous media. Materials Today Communications. 2022; 30: 102961.
[140] Chinnathambi S, Shirahata N. Recent advances on fluorescent biomarkers of near-infrared quantum dots for *in vitro* and *in vivo* imaging. Science and Technology of Advanced Materials. 2019; 20: 337–355.
[141] Fan Z, Li S, Yuan F, Fan L. Fluorescent graphene quantum dots for biosensing and bioimaging. RSC Advances. 2015; 5: 19773–19789.
[142] Lu H, Li W, Dong H, Wei M. Graphene Quantum Dots for Optical Bioimaging. Small (Weinheim an Der Bergstrasse, Germany). 2019; 15: e1902136.
[143] Pons T, Bouccara S, Loriette V, Lequeux N, Pezet S, Fragola A. In Vivo Imaging of Single Tumor Cells in Fast-Flowing Bloodstream Using Near-Infrared Quantum Dots and Time-Gated Imaging. ACS Nano. 2019; 13: 3125–3131.

[144] Cihalova K, Hegerova D, Jimenez AM, Milosavljevic V, Kudr J, Skalickova S, et al. Antibody-free detection of infectious bacteria using quantum dots-based barcode assay. Journal of Pharmaceutical and Biomedical Analysis. 2017; 134: 325–332.
[145] Qiu L, Cui P, Zhu Z, Xu M, Jia W, Sheng J, et al. Multienzyme detection and in-situ monitoring of enzyme activity by bending CE using quantum dots-based polypeptide substrate. Electrophoresis. 2020; 41: 1103–1108.
[146] Jiang XL, Liu JH, Que YT, Que YM, Hu PP, Huang CZ, et al. Multifunctional Single-Layered Graphene Quantum Dots Used for Diagnosis of Mitochondrial Malfunction-Related Diseases. ACS Biomaterials Science & Engineering. 2020; 6: 1727–1734.
[147] Liang Y, Zhang T, Tang M. Toxicity of quantum dots on target organs and immune system. Journal of Applied Toxicology: JAT. 2022; 42: 17–40.
[148] Reshma VG, Mohanan PV. Quantum dots: Applications and safety consequences. Journal of Luminescence. 2019; 205: 287–298.
[149] Sedighi M, Mahmoudi Z, Abbaszadeh S, Eskandari MR, Saeinasab M, Sefat F. Nanomedicines for hepatocellular carcinoma therapy: Challenges and clinical applications. Materials Today Communications. 2023; 34: 105242.
[150] Lu J, Tang M, Zhang T. Review of toxicological effect of quantum dots on the liver. Journal of Applied Toxicology: JAT. 2019; 39: 72–86.
[151] Wu D, Lu J, Ma Y, Cao Y, Zhang T. Mitochondrial dynamics and mitophagy involved in MPA-capped CdTe quantum dots-induced toxicity in the human liver carcinoma (HepG2) cell line. Environmental Pollution (Barking, Essex: 1987). 2021; 274: 115681.
[152] Wu T, Tang M. Toxicity of quantum dots on respiratory system. Inhalation Toxicology. 2014; 26: 128–139.
[153] Yang L, Kuang H, Zhang W, Wei H, Xu H. Quantum dots cause acute systemic toxicity in lactating rats and growth restriction of offspring. Nanoscale. 2018; 10: 11564–11577.
[154] Stern ST, Zolnik BS, McLeland CB, Clogston J, Zheng J, McNeil SE. Induction of autophagy in porcine kidney cells by quantum dots: a common cellular response to nanomaterials? Toxicological Sciences: an Official Journal of the Society of Toxicology. 2008; 106: 140–152.
[155] Zhang M, Bishop BP, Thompson NL, Hildahl K, Dang B, Mironchuk O, et al. Quantum Dot Cellular Uptake and Toxicity in the Developing Brain: Implications for Use as Imaging Probes. Nanoscale Advances. 2019; 1: 3424–3442.
[156] Levy M, Chowdhury PP, Nagpal P. Quantum dot therapeutics: a new class of radical therapies. Journal of Biological Engineering. 2019; 13: 48.
[157] Wang X, Tian J, Yong KT, Zhu X, Lin MCM, Jiang W, et al. Immunotoxicity assessment of CdSe/ZnS quantum dots in macrophages, lymphocytes and BALB/c mice. Journal of Nanobiotechnology. 2016; 14: 10.
[158] Wang ZG, Zhou R, Jiang D, Song JE, Xu Q, Si J, et al. Toxicity of Graphene Quantum Dots in Zebrafish Embryo. Biomedical and Environmental Sciences: BES. 2015; 28: 341–351.
[159] Amiri G, Valipoor A, Parivar K, Modaresi M, Noori A, Gharamaleki H, et al. Comparison of Toxicity of CdSe: ZnS Quantum Dots on Male Reproductive System in Different Stages of Development in Mice. International Journal of Fertility & Sterility. 2016; 9: 512–520.
[160] Xu G, Lin G, Lin S, Wu N, Deng Y, Feng G, et al. The Reproductive Toxicity of CdSe/ZnS Quantum Dots on the in vivo Ovarian Function and in vitro Fertilization. Scientific Reports. 2016; 6: 37677.
[161] Kang YF, Li YH, Fang YW, Xu Y, Wei XM, Yin XB. Carbon Quantum Dots for Zebrafish Fluorescence Imaging. Scientific Reports. 2015; 5: 11835.
[162] Mahendra S, Zhu H, Colvin VL, Alvarez PJ. Quantum dot weathering results in microbial toxicity. Environmental Science & Technology. 2008; 42: 9424–9430.
[163] Chakraborty D, Ethiraj KR, Chandrasekaran N, Mukherjee A. Mitigating the toxic effects of CdSe quantum dots towards freshwater alga Scenedesmus obliquus: Role of eco-corona. Environmental Pollution (Barking, Essex: 1987). 2021; 270: 116049.
[164] Chandra A, Deshpande S, Shinde DB, Pillai VK, Singh N. Mitigating the Cytotoxicity of Graphene Quantum Dots and Enhancing Their Applications in Bioimaging and Drug Delivery. ACS Macro Letters. 2014; 3: 1064–1068.
[165] Romoser A, Ritter D, Majitha R, Meissner KE, McShane M, Sayes CM. Mitigation of quantum dot cytotoxicity by microencapsulation. PloS One. 2011; 6: e22079.
[166] Ramachandran P, Lee CY, Doong RA, Oon CE, Kim Thanh NT, Lee HL. A titanium dioxide/nitrogen-doped graphene quantum dot nanocomposite to mitigate cytotoxicity: synthesis, characterisation, and cell viability evaluation. RSC Advances. 2020; 10: 21795–21805.
[167] Khan MS, Sheikh A, Abourehab MAS, Gupta N, Kesharwani P. Understanding the theranostic potential of quantum dots in cancer management. Materials Today Communications. 2023; 36: 106424.
[168] Hoshino A, Hanaki KI, Suzuki K, Yamamoto K. Applications of T-lymphoma labeled with fluorescent quantum dots to cell tracing markers in mouse body. Biochemical and Biophysical Research Communications. 2004; 314: 46–53.
[169] Tang Y, Han S, Liu H, Chen X, Huang L, Li X, et al. The role of surface chemistry in determining in vivo biodistribution and toxicity of CdSe/ZnS core-shell quantum dots. Biomaterials. 2013; 34: 8741–8755.
[170] Lovrić J, Bazzi HS, Cuie Y, Fortin GRA, Winnik FM, Maysinger D. Differences in subcellular distribution and toxicity of green and red emitting CdTe quantum dots. Journal of Molecular Medicine (Berlin, Germany). 2005; 83: 377–385.
[171] Wagner AM, Knipe JM, Orive G, Peppas NA. Quantum dots in biomedical applications. Acta Biomaterialia. 2019; 94: 44–63.
[172] Cinteza LO. Quantum dots in biomedical applications: advances and challenges. J Nanophotonics. 2010; 4: 042503.
[173] Pang C, Gong Y. Current Status and Future Prospects of Semiconductor Quantum Dots in Botany. Journal of Agricultural and Food Chemistry. 2019; 67: 7561–7568.
[174] Barve K, Singh U, Yadav P, Bhatia D. Carbon-based designer and programmable fluorescent quantum dots for targeted biological and biomedical applications. Materials Chemistry Frontiers. 2023; 7: 1781–802.
[175] Henna TK, Pramod K. Graphene quantum dots redefine nanobiomedicine. Materials Science & Engineering. C, Materials for Biological Applications. 2020; 110: 110651.
[176] Miao Y, Wang X, Sun J, Yan Z. Recent advances in the biomedical applications of black phosphorus quantum dots. Nanoscale Advances. 2021; 3: 1532–1550.